\documentclass[12pt]{iopart}

\usepackage{iopams}  
\usepackage{graphicx}
\usepackage{subfigure}
\usepackage{cite}
\begin{document}

\title[Article Title]{Unraveling effects of electron correlation in two-dimensional Fe$_{n}$GeTe$_{2}$ (n=3, 4, 5) by dynamical mean field theory}

\author{Sukanya Ghosh, Soheil Ershadrad, Vladislav Borisov and Biplab Sanyal*}

\address{Department of Physics and Astronomy, Uppsala University, Box-516, 75120, Uppsala, Sweden}
\ead{*biplab.sanyal@physics.uu.se}
\vspace{10pt}

\begin{abstract}
The Fe$_{n}$GeTe$_{2}$ systems are newly discovered  two-dimensional van-der-Waals materials, exhibiting magnetism at room temperature. The sub-systems belonging to Fe$_{n}$GeTe$_{2}$ class are special because they show site-dependent magnetic behavior. We focus on the critical evaluation of magnetic properties and electron correlation effects in Fe$_{n}$GeTe$_{2}$ ($n$= 3, 4, 5) (FGT) systems performing first-principles calculations. Three different ab-initio approaches have been used, viz., i) standard density functional theory (DFT), ii) incorporating static electron correlation (DFT+U) and iii) inclusion of dynamic electron correlation effect (DFT+DMFT). Our results show that DFT+DMFT is the most accurate technique to correctly reproduce the magnetic interactions and experimentally observed transition temperatures. The inaccurate values of structural parameters, magnetic moments and exchange interactions obtained from DFT+U make this method inapplicable for the FGT family. Correct determination of magnetic properties for this class of materials is important since they are promising candidates for spin transport and spintronic applications at room temperature.

\end{abstract}




\maketitle









\section{Introduction}\label{sec1}
Magnetism and spintronics have been the topic of interest for fundamental studies as well as technological applications. Two-dimensional (2D) van der Waals (vdW) materials have recently developed increasing attention, as they show a new direction to explore magnetism in low-dimension. Since the discovery of graphene\cite{Novoselov_2007}, the world of two-dimensional (2D) materials is rapidly expanding to an enormous variety of systems due to their interesting properties. The discovery of intrinsic long-range magnetic order is one of the most exciting recent developments in 2D materials\cite{Nat_rev_phys_2019,Nat_fm_2017}. The newly discovered 2D van-der-Waals (vdW) materials offer a new means to study magnetism in low-dimension, where spin fluctuations are expected to be strongly enhanced at finite temperature destroying long-range magnetic order according to Mermin-Wagner theorem\cite{mermin1966absence}. Magnetic anisotropy reduces spin fluctuations and results in finite Curie temperature below which a magnetic order in 2D can survive and thereby lifting the restrictions imposed by Mermin-Wagner theorem\cite{Nature_2020,CrI3_npj}. Magnetism in 2D has started a new era in the field of energy-efficient device fabrication due to its tunability under small or moderate-scale external perturbations\cite{Adv_Mater_2020}.

The 2D magnets play a crucial role in the development of efficient spintronic devices \cite{vzutic2004spintronics} not only by size reduction but also by introducing novel physics stemming from 2D confinement.
Prior to 2016, doping by magnetic impurities was the main strategy to induce magnetism in 2D materials, which has failed to materialize 2D magnets with high Curie temperature \cite{cortie2020two}. The exfoliation of 2D insulating ferromagnet CrI$_{3}$\cite{gong2017discovery} and Cr$_{2}$Ge$_{2}$Te$_{6}$\cite{huang2017layer}, in 2016, opened the doors to the application of 2D magnets in semiconducting devices. 

However, the relatively low transition temperature ($T_\mathrm{C} \approx60 K$) intrinsic to the 2D magnetic insulators and/or semiconductors limits their utilization. On the other hand, metals exhibiting 2D magnetism are more promising for practical applications due to their high transition temperature. The most significant advantage of the metallic ferromagnets is that their conducting nature enables an interplay between spin and charge degrees of freedom, which lies at the heart of spintronics\cite{BHATTI2017530,Tsymbal_2019,Tsymbal_2022}. There are recent studies on ferromagnetic metals, such as CrTe$_{x}$, Cr$_{2}$BC, FeSe$_{2}$, FeTe, MnSe, and Fe$_{n}$GeTe$_{2}$, reporting high $T_\mathrm{C}$ (130–846K)\cite{Zhang_Adv_Matr_2019,FeTe_2020,Abdullahi_2021,Wu_2021,Meng_2021}. The newly discovered 2D metallic Fe–Ge–Te ternary compounds or popularly known as `FGT' systems exhibit $T_\mathrm{C}$ close to room temperature and gain tremendous attention due to their interesting magnetic properties. 

Each member belonging to the FGT family exhibits a plethora of interesting physical properties, and among them, Fe$_{5}$GeTe$_{2}$ is the most complicated system. Recent experimental studies have found anomalous magnetic behavior in bulk Fe$_{5}$GeTe$_{2}$ at temperature $<$ 100 K, which is speculated as a result of magnetostructural effects\cite{Ramesh_PRB_2020,may2019ferromagnetism,Ly_Adv_Mater_2021}. Recently, Liu {\textit{et al}}, using the DFT+U approach, proposed that a competition between ferromagnetic (FM) and antiferromagnetic (AFM) coupling can give rise to a transition below 200 K in Fe$_{5}$GeTe$_{2}$, and they show the net magnetization reduces with a decrease in temperature similar to the experimental reports. However, they failed to capture the traces of this transition in Fe$_{4}$GeTe$_{2}$ \cite{liu2022layer}. Though the study by Liu {\textit{et al}} can reproduce the magnetization vs. temperature behavior of Fe$_{5}$GeTe$_{2}$, as observed in experiments, the moments of Fe atoms reported in their study are hugely overestimated (by $\sim$ 1 $\mu_{B}$). This suggests that the DFT+U technique cannot correctly predict the magnetic moments of Fe sublattices present in FGT systems. Similarly, other first principles studies also revealed that DFT+U is not a good approximation for the FGT systems to produce the correct lattice parameters and magnetic moments \cite{JOE2019299,F3GT_DFT_2016}. 

Electrons occupying the $s$ and $p$ orbitals are strongly itinerant and their kinetic energies are dominant over the Coulomb repulsion. Therefore, a static mean field approximation (like DFT) should be suitable enough to study these weakly correlated electrons. On the other hand, due to significant Coulomb interactions between electrons in the $d$ and $f$ orbitals, strongly correlated electrons become localized on their atomic sites. In this regime, DFT is no longer a proper approach and can yield inaccurate results. The possible solutions are DFT+U in the static mean field approximation, mostly applicable for Mott insulators and dynamical mean field theory with frequency-dependent self-energy for strongly correlated metals. 
The presence of itinerant electrons is the cause of spontaneous magnetism in 2D metals, which can be explained by the Stoner model\cite{Stoner}. However, in the case of metallic FGT systems, some studies show the evidence of non-Stoner magnetism where the local moments may play crucial role in the electronic and magnetic properties\cite{F3GT_JPCL,Non_Stoner}. These results hint towards the possibility that the FGT systems could be an admixture of localized and itinerant electrons, exhibiting moderate electron correlation. Hence, the use of DMFT will be more justified.



The main peculiarity related to all the FGT systems is their site-dependent magnetic and electronic properties\cite{may2019ferromagnetism}. In other words, based on the magnetic moments associated with different Fe sublattices, electrons belonging to some Fe atoms are expected to have more itinerant character than others. All these complications combined together have made the proper treatment of electronic structure and magnetic behavior in FGT systems a controversial issue. Up to this date, researchers have used DFT\cite{Blugel_2022,JOE2019299,F3GT_DFT_2016}, DFT+U\cite{liu2022layer} or DMFT\cite{Sassa_DMFT,Sci_adv_FGT_family} approaches, leaving behind the most crucial question unanswered i.e. which method is the most reliable one to correctly treat the FGT systems. In this paper, we perform a systematic study, comparing different aspects of electronic and magnetic properties of Fe$_{3}$GeTe$_{2}$, Fe$_{4}$GeTe$_{2}$, and Fe$_{5}$GeTe$_{2}$, mostly in their 2D form (except Fe$_{3}$GeTe$_{2}$), via DFT, DFT+U, and DFT+DMFT approaches. We put forward a detailed comparative study to analyse the outcomes of these different computational techniques and their compatibility with experimental results. Since atomically thin films or monolayers of FGT systems are not experimentally studied yet, we considered the well-studied bulk Fe$_{3}$GeTe$_{2}$ as our reference to examine the accuracy of our calculations, and bridge between 3D and 2D regimes. 



\section{Results and Discussion}\label{sec2}

\subsection{Systems}\label{sec21}

\begin{figure*} [t]
\includegraphics[width=0.95\textwidth] {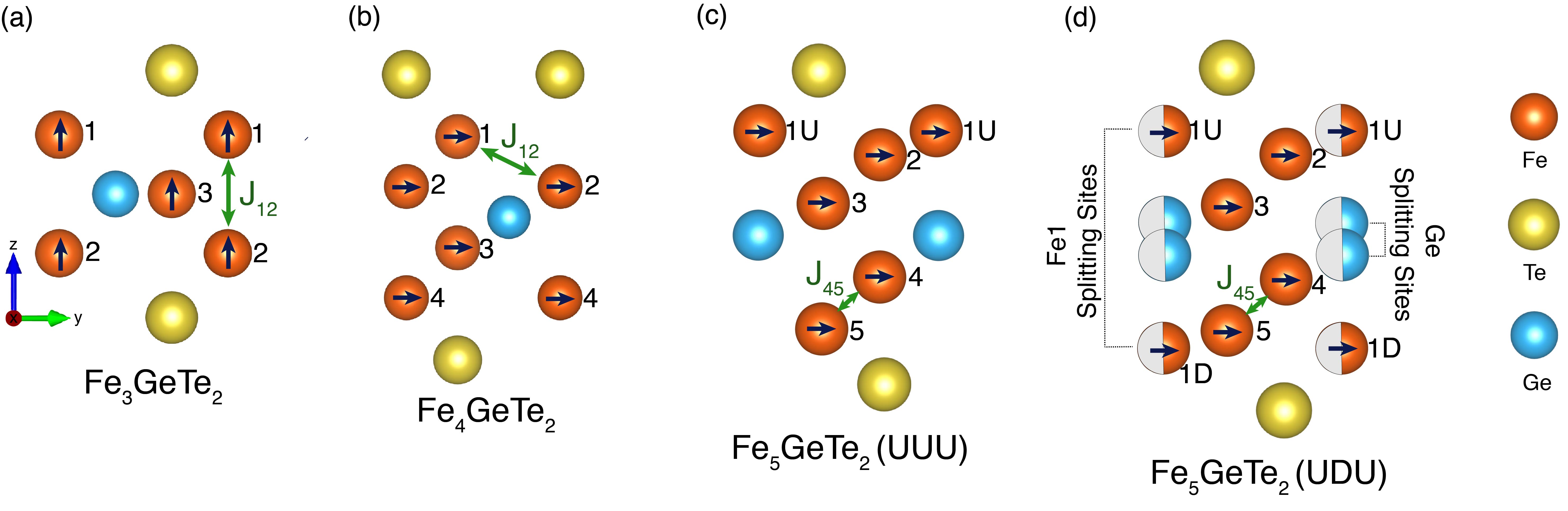}
\centering
\caption{The schematic side views of (a) Fe$_{3}$GeTe$_{2}$, (b) Fe$_{4}$GeTe$_{2}$, (c) UUU Fe$_{5}$GeTe$_{2}$, and (d) UDU Fe$_{5}$GeTe$_{2}$ monolayers, where orange, yellow, and blue circles represent Fe, Te, and Ge atoms, respectively. Direction of the arrows shown inside Fe atoms, indicates the easy axis of magnetization obtained from our DFT+DMFT calculations. The half-colored circles in Fig.~\ref{fig1}(d) show the Fe1-Ge split sites present in the UDU configuration. The green arrow drawn in each system shows the pair of Fe atoms exhibiting the strongest isotropic symmetric exchange interaction between each other.}
\label{fig1}
\end{figure*}

    Here we focus on the Fe$_{n}$GeTe$_{2}$ monolayers, where $n=3, 4$ and 5, to investigate the magnetic properties of FGT systems using different first-principles techniques, e.g., standard density functional theory (DFT), DFT with the inclusion of dynamic and static electron correlation effects.

    In this section, we discuss the structural properties of the monolayers. In the case of  Fe$_{5}$GeTe$_{2}$, we study both without and with the Fe1-Ge split sites, see Ref~\citenum{SE_jpcl} for details. Fig.~\ref{fig1} shows the side view of the Fe$_{n}$GeTe$_{2}$ monolayers, (a) $n=3$, (b) $n=4$, (c) $n=5$ in UUU and (d) UDU configurations, we label each Fe atom present in the unit cell of Fe$_{n}$GeTe$_{2}$ monolayers. The space-groups (and in-plane lattice parameters) for Fe$_{3}$GeTe$_{2}$, Fe$_{4}$GeTe$_{2}$, Fe$_{5}$GeTe$_{2}$ UUU and UDU monolayers are P$\overline{6}m$2 ($a$=4.05 \AA), P$\overline{3}m1$ ($a$= 3.97 \AA), P$3m1$ ($a$= 3.99 \AA) and P$31m$  ($a$=7.00\AA), respectively. From the type of space-groups possessed by Fe$_{n}$GeTe$_{2}$ monolayers, it is clear that Fe$_{5}$GeTe$_{2}$ possesses lower symmetry operations, and hence less symmetric compared to $n$ =3 and 4.

    Each Fe$_{n}$GeTe$_{2}$ system contains different Fe sublattices based on their structural arrangements. This sublattice classification is reflected on the electronic and magnetic properties \cite{Kim_2021,SE_jpcl}. In the case of Fe$_{3}$GeTe$_{2}$, Fe1 and Fe2 sublattices are equivalent, while Fe3 is different. Similarly, for Fe$_{4}$GeTe$_{2}$, Fe1 and Fe4 species belong to the same category, whereas Fe2 and Fe3 are similar. 
    
    The structural configuration for Fe$_{5}$GeTe$_{2}$ is more complex. Based on the previous experimental and theoretical studies, the Fe1 species can occupy two possible sites either above or below Ge, giving rise to Fe1-Ge split-sites\cite{Ramesh_PRB_2020,may2019ferromagnetism,SE_jpcl}. Also, depending on the thermal history during the synthesis process, this system may show the absence of Fe1-Ge splitting\cite{may2020tuning,may2019ferromagnetism}. To incorporate the Fe1-Ge splitting, a $\sqrt{3}\times\sqrt{3}$ cell of Fe$_{5}$GeTe$_{2}$ monolayer is constructed, where two (one) Fe1 are (is) situated above (below) Ge, this structure is referred to as `UDU' configuration. In the absence of Fe1-Ge splitting, the Fe1 species is placed at top of the Ge atom, this structure is named as `UUU' configuration. From our previous first-principles study, we found that the UDU configuration is energetically favored over the UUU configuration\cite{SE_jpcl}, resulting in the $\sqrt{3}\times\sqrt{3}$ superstructure, as already discussed by previous experimental studies\cite{SE_jpcl,Ramesh_PRB_2020,may2020tuning,may2019ferromagnetism,Ly_Adv_Mater_2021}. Hence, in the following sections, we mainly focus on the UDU configuration of Fe$_{5}$GeTe$_{2}$ monolayer.
    All the five Fe atoms present in the unit cell of UUU configuration are different, hence this configuration is the least symmetric. In the case of UDU, we find the Fe atoms can be roughly (not strictly) categorized into four sublattices in UDU configuration: i) Fe2, Fe5, ii) Fe3, Fe4, iii) Fe1U and iv) Fe1D.  The sublattice classification in Fe$_{n}$GeTe$_{2}$ systems has a direct consequence on the exchange interactions present in these systems and also the magnetization vs. temperature behavior as reported in Ref.~\citenum{SE_jpcl}. The black arrows drawn in Figs.~\ref{fig1}(a)-(d) show the direction of the easy axis of magnetization for different Fe$_{n}$GeTe$_{2}$ monolayers obtained from our DMFT studies, which we discuss in section~\ref{sec23}.

    From Fig.~\ref{fig1}, one can see that the chemical environment, coordination number and relative position of each Fe atom w.r.t. the other Fe and Ge atoms present in the Fe$_{n}$GeTe$_{2}$ monolayers change with $n$. For example, in the case of Fe$_{3}$GeTe$_{2}$, the Fe1 and Fe2 atoms are situated symmetrically relative to the Ge atom with Fe1(Fe2)-Ge distance of 2.65 \AA. While the Fe3 species is placed in the same $z$ plane as Ge with a distance 2.34 \AA\ along the $xy$-plane. The first nearest neighbor (NN) Fe1-Fe2 and Fe1-Fe3 distances are 2.47 and 2.65 \AA, respectively. Due to this small first NN distance between Fe1 and Fe2, the strongest isotropic symmetric exchange interaction exists between the first NN of these two species, shown by the green arrow in Fig.~\ref{fig1}(a). The scenario is quite different for Fe$_{4}$GeTe$_{2}$, where the Ge atom is placed between Fe2 and Fe3 (shifted along the diagonal direction). The first NN Fe1-Fe2 (Fe4-Fe3) and Fe1-Fe3 (Fe4-Fe2) distances are 2.55 \AA\ and 2.49 \AA, respectively. The first NN Fe2-Ge (or Fe3-Ge) and Fe1-Ge (or Fe4-Ge) distances are 2.40 \AA\ and 2.92 \AA, respectively. Although the shortest first NN distance exists between Fe1 and Fe3 (Fe4 and Fe2), the highest isotropic symmetric exchange interaction takes place between Fe1 and Fe2, instead of between Fe1 and Fe3. The position of Ge atom relative to the Fe atoms is responsible for such an unexpected feature. In the case of Fe$_{5}$GeTe$_{2}$ monolayer in both UUU and UDU configurations, the most significant symmetric exchange interactions take place between the first NN Fe5 and Fe4. The exchange interactions present in different FGT monolayers are discussed in more detail in section~\ref{sec241}.


\subsection{Effective Hubbard parameters}\label{sec22}
\begingroup
\begin{table*}[t]
\begin{center}
{
\caption{Effective Hubbard $U$ parameter for Fe atoms in Fe$_{n}$GeTe$_{2}$ systems obtained from constrained linear response technique. }
\centering
\begin{tabular}{|c|c|c|}
\hline
System & Atom & $U_\mathrm{eff}$ (eV) \\ 
\hline
                 & Fe1 &  4.8    \\
Fe$_{3}$GeTe$_{2}$ &  Fe2 &  4.8    \\ 
                  &  Fe3  & 3.9  \\ 

 \hline
     & Fe1 &   4.5   \\
Fe$_{4}$GeTe$_{2}$ & Fe2 &  4.0    \\ 
             & Fe3  & 4.0  \\ 
           & Fe4  & 4.5  \\ 
\hline           
   & Fe1U &     3.7   \\
   &    Fe1D  &    4.0  \\ 
Fe$_{5}$GeTe$_{2}$   &  Fe2 &   4.6   \\
  (UDU) &    Fe3  &  4.1   \\ 
   &  Fe4  &  4. 1    \\
  & Fe5  &   4.6   \\
 \hline 

\end{tabular}
\label{U}
}
\end{center}
\end{table*}
\endgroup

      As already mentioned in section~\ref{sec21}, the FGT systems possess site-dependent electronic and magnetic properties. In our previous first-principles study on Fe$_{5}$GeTe$_{2}$ monolayer, we have found that though the structural properties and magnetic moments obtained using the standard DFT technique are in good agreement with the experimental reports, the calculated Curie temperature (using Monte Carlo simulations solving the Spin-Hamiltonian) is almost twice of the experimental value reported for this system\cite{SE_jpcl}. This hints toward the fact that one needs to go beyond the standard DFT formalism to correctly capture the magnetic interactions. One possibility to solve such discrepancy in Curie temperature is to include the electron correlation effect in first-principles calculations which might be able to reproduce the experimental results. Here, we consider both the dynamic and static electron correlation effects and finally compare the results obtained using DFT, DFT+DMFT and DFT+U techniques.

    As the choice of Hubbard $U$ parameter is always ambiguous to some extent, we have calculated the effective Hubbard $U$ parameter ($U_\mathrm{eff}$) for each Fe species in its magnetic ground state (as obtained from standard DFT) using the constrained linear response (cLR) method proposed by Cococcioni {\textit{et al}}\cite{CLR}. This method is focused on the main effect associated with the on-site Coulomb repulsion $U$, neglecting the secondary effects related to the higher-multipolar terms in the Coulomb interaction. The effects of Hund's exchange interaction ($J_\mathrm{H}$) are incorporated by redefining the Hubbard $U$ as $U_\mathrm{eff}=U-J_\mathrm{H}$ (see section~\ref{sec45} for further details). The site-dependent $U_\mathrm{eff}$ values obtained using cLR for each Fe species belonging to different FGT monolayers are reported in Table~\ref{U}.

    There are previous reports in literature where the Fe$_{n}$GeTe$_{2}$ systems are studied considering (mostly) the static electron correlation effect, where the site-dependence of Hubbard $U$ is not considered\cite{Sci_adv_FGT_family,F3GT_PRB_2021}, except by Zhu {\textit{et al.}}, where $U$= 5.5 eV and 5.0 eV with the common $J_\mathrm{H}=$0.8 eV are considered for two different Fe sublattices present in Fe$_{3}$GeTe$_{2}$\cite{Sassa_DMFT}. To the best of our knowledge, apart from our work, so far, there is no other first-principles study reporting site-dependent $U$ value computed from either constrained linear response or any other standard computational technique (e.g., constrained random phase approximation or cRPA) generally followed for such calculations. It should be noted that the average value of $U_\mathrm{eff}$ considered by the previous authors to study any Fe$_{n}$GeTe$_{2}$ system falls in the range from 4 to 4.5 eV\cite{Sassa_DMFT,Sci_adv_FGT_family,F3GT_PRB_2021}, which is close to our calculated $U_\mathrm{eff}^\mathrm{avg}$ values for the Fe$_{n}$GeTe$_{2}$ monolayers. Nevertheless, it is worth mentioning that Fe$_{5}$GeTe$_{2}$ shows a much larger variation of site-dependent U values.

     It is known that the size of the magnetic moment is inversely proportional to the width of the electronic density of states. The width of electronic density of states of Fe$_{n}$GeTe$_{2}$ system is $\sim$ 8 eV (see Figs.~S1 and S2 in SI), and with $U_\mathrm{eff}\sim$ 4 eV, $U/W \sim 0.5$. Therefore, due to the existence of broad enough Fe-$d$ bands, the Fe-$d$ electrons are quite delocalized and the applicability of Hubbard U correction within the DFT+U formalism is questionable.

     In Fe$_{n}$GeTe$_{2}$ systems, the Fe sublattices can be classified into  different categories based on their magnetic properties. For example, Fe$_{3}$GeTe$_{2}$ contains two types of Fe sublattices with different magnetic moments\cite{May_F3GT_2016,Sassa_DMFT,Inorg_F3GT_2015}, e.g., 2.18 $\mu_{B}$ and 1.80 $\mu_{B}$ based on the data obtained from neutron powder diffraction experiments \cite{May_F3GT_2016}. From our DFT calculations, we find the moment of two Fe sublattices present in Fe$_{3}$GeTe$_{2}$ to be 2.49 $\mu_{B}$ and 1.50 $\mu_{B}$. This implies that the electronic nature or localization of Fe-$d$ states must be different for these two sublattices (see Fig.~S1 in SI). Using the cLR method we find the Fe species with larger moment has $U_\mathrm{eff}$=4.8 eV, while for the other Fe sublattice $U_\mathrm{eff}$=3.9 eV, supporting the relationship between the degree of localization of electronic state and magnetic moment. Similarly, Fe${4}$GeTe$_{2}$ and Fe${5}$GeTe$_{2}$ systems also exhibit site dependent magnetic properties\cite{may2019ferromagnetism,Sci_adv_FGT_family}, and this feature is reflected in our calculated $U_\mathrm{eff}$, see Table~\ref{U}.
    

\subsection{Role of electron correlation effect: comparison between DMFT, DFT and DFT+U approaches}\label{sec23}

\begingroup
\begin{table*}[h]
{
\caption{Magnetic moments associated with each Fe atom in Fe$_{n}$GeTe$_{2}$ monolayer (computed with different techniques).  }
\begin{tabular}{|c|c|c|c|c|c|}
\hline
System & species & DMFT ($\mu_{B}$)  & DFT ($\mu_{B}$) & DFT+U ($\mu_{B}$) & Expt. ($\mu_{B}$) \\ 
\hline
Fe$_{3}$GeTe$_{2}$  &  Fe1  &  2.52 &  2.49 &  3.19 & 2.18\cite{May_F3GT_2016}, 1.95\cite{Inorg_F3GT_2015} \\
                    &  Fe2 & 2.52   &  2.49  & 3.19 & 2.18\cite{May_F3GT_2016}, 1.95\cite{Inorg_F3GT_2015} \\ 
                    &  Fe3  & 1.31  & 1.50  & 2.50 & 1.54\cite{May_F3GT_2016}, 1.56\cite{Inorg_F3GT_2015} \\ 
                    & average & 2.12  & 2.16  &  2.96 & 1.60\cite{Japan_F3GT_2013}, 1.70\cite{Sassa_DMFT} \\
\hline
 Fe$_{4}$GeTe$_{2}$ & Fe1 & 2.61  & 2.57  & 3.33  &  \\
                    & Fe2 &  1.58 & 1.75  & 2.62  &  \\
                    & Fe3 &  1.58 & 1.75  & 2.62 &   \\
                    & Fe4 & 2.61  & 2.57  & 3.33  &   \\
                    & average & 2.10 & 2.16 & 2.98 & 1.80\cite{Sci_adv_FGT_family} \\
\hline
Fe$_{5}$GeTe$_{2}$ & Fe1U &  -0.45 & 1.230 & 2.77  &  ranges from\\
                   & Fe1D & 1.76 & 1.713 & 3.00 & 0.8 to 2.5\cite{may2019ferromagnetism}  \\
      (UDU)           & Fe2 & 2.55 & 2.511 & 2.97  &  \\
                    & Fe3 & 1.92  & 2.084 & 2.85 &   \\
                    & Fe4 & 1.85  & 1.979 & 2.64  &   \\ 
                    & Fe5 &  2.63 & 2.567 & 2.97  &   \\  
                    & average &  1.85  &  2.11 & 2.85 &  1.95\cite{Riberio_npj2020}, 1.80\cite{Nature_F3GT_2018} \\
 \hline

\hline   

\end{tabular}
\label{Moment}
}
\end{table*}
\endgroup

Within the framework of standard density functional theory, the exchange-correlation functionals (LDA and GGA) are derived in the limit of a nearly uniform electron density, which can not correctly describe the electronic and magnetic properties of localized electronic states with strong Coulomb interactions. In some systems, the on-site Coulomb interaction ($U$) is quite significant and becomes much larger than the band width ($W$). When $U/W<<$ 1 the system is weakly correlated and well described by LDA/GGA. In the case of strongly correlated regime $U/W >>$ 1, which gives atomic-like behavior (e.g., Hubbard bands, Mott insulators). The most interesting case is when $U/W \sim$ 1, which is a mixture of both band-like and atomic-like behavior, this can give rise to exotic phenomena like Kondo behavior, materials with heavy fermions etc.

DFT+U deals with on-site Coulomb interaction which is applicable only in the limit of $U/W >>$ 1. While the dynamical mean field or DMFT covers the entire range of the parameter $U/W$, i.e., from uncorrelated metal to Mott insulator. Since the Fe$_{n}$GeTe$_{2}$ systems are metallic in nature, inclusion of static electron correlation or DFT+U is not supposed to be a good choice. To check the applicability of DMFT vs. DFT+U we calculate and compare the magnetic moments, exchange interactions, magnetic anisotropies and finally the Curie temperatures.


The concept of dynamical mean-field theory is to replace a lattice model with many degrees of freedom by an effective single-site model coupled to a self-consistent bath\cite{Georges_DMFT}. The primary quantity of DMFT is the local Green's function $G_{R}(z)$, defined as the one-electron Green's function projected to the correlated states at site $R$: $G_{R}(z)=P_{R}G(z)P_{R}$, where $P_{R}$ is the projection operator. $G(z)$ is the one-electron Green's function, which is given by:
\begin{equation}
    G(z)=[z+\mu-H_\mathrm{eff}-\Sigma(z)]^{-1}, 
    \label{G}
\end{equation}
where $\mu$ is the chemical potential, the term $H_\mathrm{eff}$ is the effective Hamiltonian which includes the Hartree, exchange and correlation terms on the level of the generalized-gradient approximation (GGA). $\Sigma(z)$ is the self-energy, which goes beyond GGA and includes dynamical electronic correlations and $z=i\omega_{n}$ where $\omega_{n}$ is the Matsubara frequency. The approximation of DMFT assumes the locality of one-electron self-energy, which can be written as the sum of local self-energies for all Bravais lattice sites. In DMFT, the problem of a lattice is substituted by the impurity problem, popularly known as the effective Anderson model, where a single correlated site in the self-consistent bath is described by the bath Green's function $\mathcal{G}_{0}(R,z)$ defined as:
\begin{equation}
\mathcal{G}_{0}^{-1}(R,z)=G_{R}^{-1}(z)+\Sigma_{R}(z)
\label{DMFT}
\end{equation}


Here, we perform fully charge self-consistent DFT combined with DMFT\cite{Georges_DMFT} by spin-polarized T-matrix combined with fluctuating exchange approximation or SPTF solver\cite{SPTF_PRB}, which is a perturbative solver, best suited for weak correlations, computationally less expensive, applicable for spin-polarized case treating crystal field and spin–orbit effects correctly, and gives accurate DOS and spectral densities\cite{Wills2000}. 
We use site-dependent Hubbard $U_\mathrm{eff}$ parameter for DMFT and DFT+U calculations obtained from the constrained linear response method, as reported in Table~\ref{U}.

In principle, DMFT is based on a mapping of lattice models onto quantum impurity models and for correlated electrons, this mapping is exact in the limit of infinite dimensions\cite{DMFT_1996}. However, in recent days, DMFT has been proven to be a reliable and well-controlled approximation to study correlation effects in bulk solids as well as for two-dimensional systems\cite{Volodymyr_2012}. Recent studies on 2D or layered materials reported that DFT+DMFT technique describes the electronic and magnetic properties more accurately than DFT+U or standard DFT\cite{Kim_2020,Zhou_2021,Kvashnin_2022}.

First, we focus on the magnetic moments calculated using DMFT, DFT and DFT+U techniques. Table~\ref{Moment} reports the magnetic moment of each Fe sublattice for Fe$_{3}$GeTe$_{2}$, Fe$_{4}$GeTe$_{2}$ and Fe$_{5}$GeTe$_{2}$ (UDU configuration) monolayers (moments of Fe$_{5}$GeTe$_{2}$ in UUU configuration are reported in Table S1 in SI). As mentioned in section~\ref{sec21}, the influence of structural symmetry on the electronic/magnetic properties of Fe$_{n}$GeTe$_{2}$ systems is reflected in the magnetic moments reported in Table~\ref{Moment}. 
We find the values of magnetic moment obtained using DMFT and DFT are quite similar, except for the Fe1U sublattice for Fe$_{5}$GeTe$_{2}$ in UDU configuration. It is known from previous studies that this species exhibits fluctuating moment and is mainly responsible for magnetic anomaly observed in Fe$_{5}$GeTe$_{2}$ at low temperature\cite{may2019ferromagnetism,may2020tuning,Ramesh_PRB_2020,SE_jpcl}. The magnetic moments of the UUU configuration are reported in Table~S2 in Supplementary Information (SI). More interestingly, both the DFT+DMFT and DFT moments are close in magnitude to the moments obtained from neutron powder diffraction experiments, as we see in Table~\ref{Moment}. There is a small discrepancy between the experimental and DMFT/DFT results which arises because the DMFT or DFT moments are calculated at 0K whereas the neutron powder diffraction experiments are performed at finite temperatures ($\sim 2-5$ K)\cite{may2019ferromagnetism,Inorg_F3GT_2015,May_F3GT_2016}. 


\begin{figure*} [t]
\centering
\includegraphics[width=1.0\textwidth] {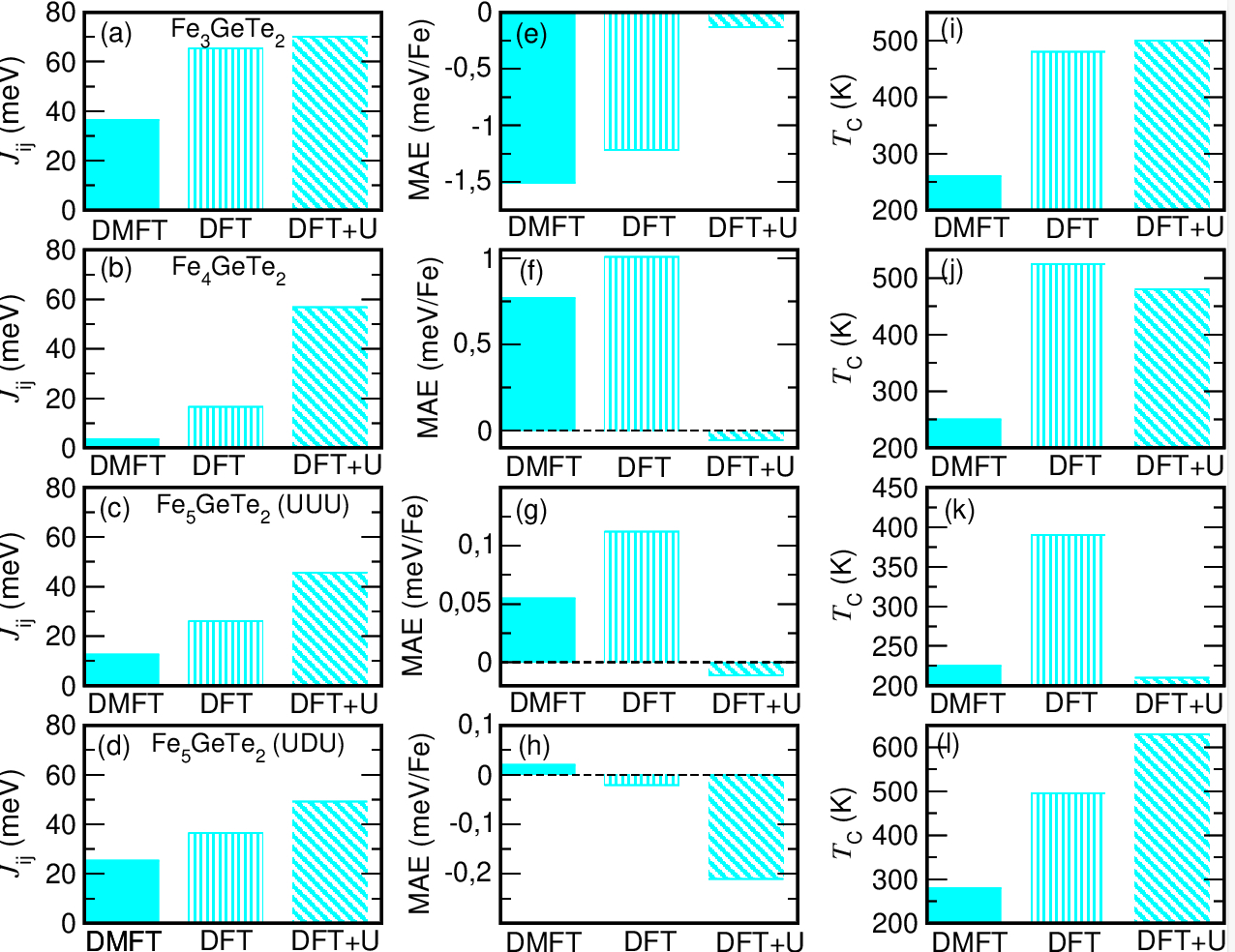}
\caption{(a)-(d) Value of the highest $J_{ij}$ interaction for (a) Fe$_{3}$GeTe$_{2}$, Fe$_{4}$GeTe$_{2}$ and Fe$_{5}$GeTe$_{2}$ (in both UUU and UDU configurations) monolayers obtained using DMFT, DFT and DFT+U. (e)-(h) Magnetic anisotropy energy (MAE=E$_{\perp}$-E$_{\parallel}$) Fe$_{3}$GeTe$_{2}$, Fe$_{4}$GeTe$_{2}$ and Fe$_{5}$GeTe$_{2}$ (in both UUU and UDU configurations) monolayers obtained using DMFT, DFT and DFT+U.(i)-(l) Curie temperature $T_\mathrm{C}$ for Fe$_{3}$GeTe$_{2}$, Fe$_{4}$GeTe$_{2}$ and Fe$_{5}$GeTe$_{2}$ (in both UUU and UDU configurations) monolayers obtained using DMFT, DFT and DFT+U.  }
\label{comparison}
\end{figure*}

In the case of DFT+U, a large deviation between the computed and experimental moments is observed. Moreover, the lattice parameters of FGT monolayers get hugely deviated from the experimental measurements. Previous DFT+U studies on FGT systems also report significant discrepancy in structural and electronic properties of FGT systems compared to experimental results\cite{F3GT_PRB_2021,Hu_ACS_2020,F3GT_DFT_2016,JOE2019299,liu2022layer}. Note, in the DFT+U calculations we use the fully-localized limit as the double counting (DC) correction to study the electronic and magnetic properties. We perform further calculations using the around mean field (AMF) as the DC correction using exactly the same site-dependent $U_\mathrm{eff}$ values. In that case also we find that the magnetic moments have hugely deviated from the experimental moments. This implies that DFT+U fails to reproduce the correct ground state properties of FGT systems.

Fe$_{n}$GeTe$_{2}$ systems are metallic with conducting electrons and therefore, to correctly describe the electronic properties and magnetic behavior of such metallic systems inclusion of DFT+U, i.e., consideration of on-site Coulomb repulsion is not the proper approach\cite{JOE2019299}. Instead, the dynamical electron correlation effect or DMFT may be a better choice to capture the correct electronic structure and magnetic behavior, which can properly describe the site-dependent magnetic behavior present in these systems\cite{Ramesh_PRB_2020,Sassa_DMFT,Sci_adv_FGT_family}. There are recent studies on Fe$_{3}$GeTe$_{2}$ which have mentioned the existence of heavy Fermion states due to intriguing interplay between localized magnetic moments and itinerant electrons\cite{Kondo_Nanolett,Sciadv_F3GT_2018}. Also, Zhang et al. have proposed the existence of itinerant ferromagnetism in Fe$_{5}$GeTe$_{2}$.\cite{Ramesh_PRB_2020} 
These reports hint toward the fact that the electronic nature of Fe-$3d$ states is sublattice specific. Such findings and prediction have motivated us to consider the electron correlation effect in FGT systems incorporating site-dependent $U_{eff}$. There are a few first-principles reports on FGT systems where the dynamical electron correlation effect is considered\cite{Sassa_DMFT,Sci_adv_FGT_family}, but those studies do not explain why DMFT should be considered as a better technique compared to other first-principles methods, for the correct description of FGT systems. In our study, we show the importance of the dynamic electron correlation effect for the FGT systems by calculating the exchange interactions, magnetic anisotropy and more importantly the Curie temperature of these systems.



Monte Carlo simulations are performed to calculate $T_\mathrm{C}$ using the following spin Hamiltonian:

\begin{equation}
    H = -\sum_{i\neq j} J_{ij}\vec e_{i}{\cdot} \vec e_{j}-\sum_{i\neq j} \vec{D}_{ij}\cdot (\vec e_{i}{\times} \vec e_{j})-\sum_{i\neq j} K_{i}(e_{i}^{z})^{2},
    \label{H}
\end{equation}

where $J_{ij}$ and $\vec{D}_{ij}$ are the symmetric and antisymmetric exchange interactions, respectively between the $i$th and $j$th sites. According to eq.~\ref{H}, positive (negative) sign of $J_{ij}$ implies ferromagnetic (antiferromagnetic) interaction. $K_{i}$ is the single-ion anisotropy for the $i$th site. We first check how the $J_{ij}$ and $K_{i}$ values obtained using DFT, DMFT and DFT+U are different from each other to influence the value of $T_\mathrm{C}$. Here we neglect the term $D_{ij}$ since we find that the magnitude of Dzyaloshinkii-Moriya interaction (DMI) are quite small compared to the $J_{ij}$ and MAE values, therefore the DMI is expected to have negligible influence on $T_\mathrm{C}$.

In Fig.~\ref{comparison} the histogram plots show (a)-(d) the strongest isotropic symmetric exchange interaction ($J_{ij}$), (e)-(h) magnetic anisotropy energy (MAE) and (i)-(l) Curie temperature ($T_\mathrm{C}$) for Fe$_{n}$GeTe$_{2}$ ($n=3, 4, 5$) systems obtained using DMFT, DFT and DFT+U techniques. From Figs.~\ref{comparison}(a)-(d) we see that the sign of the highest $J_{ij}$ interactions computed using these different methods remain unaltered but their magnitudes differ significantly. In the case of Fe$_{3}$GeTe$_{2}$ and Fe$_{5}$GeTe$_{2}$, we find the value of the strongest $J_{ij}$ interaction obtained using DMFT is almost half of the DFT value, while this difference is even more (DMFT $\sim$4 times smaller than DFT) for Fe$_{4}$GeTe$_{2}$. The strongest $J_{ij}$ value using DFT+U for $n=3, 4,$ and 5 is overestimated by 6\%, 71\%, 43\% (UUU) and 26\% (UDU), respectively, w.r.t. the standard DFT results for the corresponding systems. 

\begin{figure*} [t]
\begin{center}
\includegraphics[width=0.7\textwidth] {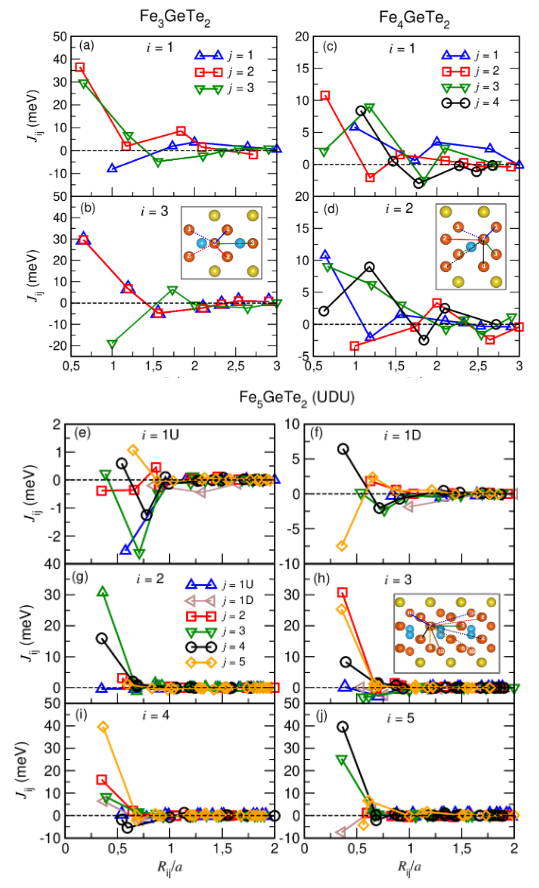}
\caption{\label{Jij_DMFT}Isotropic symmetric exchange parameters $J_{ij}$ as a function of neighbouring distance $R_{ij}$ for (a)-(b) Fe$_{3}$GeTe$_{2}$, (c)-(d) Fe$_{4}$GeTe$_{2}$ and (e)-(j) Fe$_{5}$GeTe$_{2}$ monolayer, obtained using DMFT. The $J_{ij}$ values reported here are multiplied by the corresponding coordination numbers. }
\end{center}
\label{comparison}
\end{figure*}

 Magnetic anisotropy energy MAE=$E_{\perp}-E_{\parallel}$ calculated using different computational techniques are plotted in Figs.~\ref{comparison}(e)-(h), positive (negative) sign indicates in-plane (out-of-plane) easy axis. The standard DFT results show that the easy axis of magnetization is strongly along the out-of-plane and in-plane directions for Fe$_{3}$GeTe$_{2}$ and Fe$_{5}$GeTe$_{2}$ monolayers, respectively, while for Fe$_{5}$GeTe$_{2}$ monolayer, MAE is weakly out-of-plane (in-plane) for the UDU (UUU) configuration. The DFT MAE results for Fe$_{3}$GeTe$_{2}$, Fe$_{4}$GeTe$_{2}$ and Fe$_{5}$GeTe$_{2}$ (UUU configuration) monolayers are in excellent agreement with previous DFT results\cite{Kim_2021,SE_jpcl,Yang2021,SE_jpcl}.

Next we investigate how the DFT computed MAE results get modified by the inclusion of dynamical electronic correlations. The direction of easy axis obtained using DMFT and DFT remains same for any Fe$_{n}$GeTe$_{2}$ monolayer, except for Fe$_{5}$GeTe$_{2}$ in UDU configuration. The DFT result shows weak out-of-plane anisotropy while DMFT produces small in-plane anisotropy value. The DMFT (DFT) computed MAE for Fe$_{3}$GeTe$_{2}$, Fe$_{4}$GeTe$_{2}$, Fe$_{5}$GeTe$_{2}$ in UUU and UDU configurations are: -1.51 (-1.22), 0.77 (1.01), 0.05 (0.11) and 0.02 (-0.03) meV/Fe, respectively.

It is important to note that the direction of easy axis for the Fe$_{3}$GeTe$_{2}$ and Fe$_{4}$GeTe$_{2}$ monolayers evaluated using DMFT and DFT is in good agreement with the experimental results reported for the corresponding bulk systems\cite{Sci_adv_FGT_family}. In the case of bulk Fe$_{5}$GeTe$_{2}$ the scenario is not so straightforward because some studies report weak in-plane easy axis\cite{Ramesh_PRB_2020}, while some mentions canted out-of-plane\cite{may2019ferromagnetism} configuration. This can be attributed to the complexity related to the structural properties and chemical composition of Fe$_{5}$GeTe$_{2}$. Our previous first-principles study on Fe$_{5}$GeTe$_{2}$ monolayer reported that both the direction and magnitude of MAE vary depending on the concentration of Fe vacancy and the presence of Fe1-Ge split sites\cite{SE_jpcl}. Since the strength of MAE for this system is quite weak, small structural changes or weak perturbations can easily tune this property.


Figs.~\ref{comparison} (i)-(l) show the value of $T_\mathrm{C}$ computed using different techniques for different FGT monolayers. It is quite evident from Figs.~\ref{comparison}(i)-(l) that $T_\mathrm{C}$ obtained from DFT is overestimated compared with DMFT. $T_\mathrm{C}$ obtained using DFT is almost twice of the DMFT value for Fe$_{3}$GeTe$_{2}$ and Fe$_{5}$GeTe$_{2}$, while this difference increases in the case of Fe$_{4}$GeTe$_{2}$. The $M$ vs. $T$ behavior for different Fe sublattices present in the Fe$_{n}$GeTe$_{2}$ monolayers are plotted in SI.

Carefully examining Figs.~\ref{comparison}(a)-(d) and (i)-(l), one can see there is a direct correlation between the DFT and DMFT computed $J_{ij}$ and $T_\mathrm{C}$. For example, in the case of Fe$_{3}$GeTe$_{2}$, the value of the strongest $J_{ij}$ coupling using DMFT and DFT are 36.52 meV/Fe and 65.32 meV/Fe, respectively. Similarly, $T_\mathrm{C}$ changes from 270 K (DMFT) to 480 K (DFT). This implies $T_\mathrm{C}$ is dominated by the first term $J_{ij}$ in eq.~\ref{H}. However, no such relationship between $J_{ij}$ and $T_\mathrm{C}$ is observed for DFT+U results.

It is important to note that, the value of $T_\mathrm{C}$ for Fe$_{n}$GeTe$_{2}$ monolayers obtained using DMFT fall in the same range as reported in the experiments on few-layer thick or bulk FGT systems. Since in the bulk FGT systems different formula units are stacked together via weak vdW forces, the monolayer could be considered as a good representative of the bulk, though the interlayer interactions are absent in this case. Therefore, $T_\mathrm{C}$ obtained for monolayers should be compatible with their bulk counterpart. $T_\mathrm{C}$ for monolayer (bulk) Fe$_{3}$GeTe$_{2}$, Fe$_{4}$GeTe$_{2}$ and Fe$_{5}$GeTe$_{2}$ systems obtained from DFT+DMFT (experiments) are the following: 260 K (220 K\cite{Sci_adv_FGT_family,Sassa_DMFT}), 250 K (270 K\cite{Sci_adv_FGT_family}) and 275 K (275--310 K\cite{Ramesh_PRB_2020,may2019ferromagnetism,Li_2020}), respectively. From these values of $T_\mathrm{C}$ we see there is no linear relationship as a function of $n$ for different FGT monolayers, while the $T_\mathrm{C}$ values reported in experiments show the linear trend for bulk FGT systems\cite{Ramesh_PRB_2020}. Even in the case of DFT+DMFT, there is still some overestimation of $T_\mathrm{C}$ obtained for monolayer Fe$_{3}$GeTe$_{2}$ compared to the experimental values (130 K - 150 K) reported for this system\cite{May_Nat_Mater_2018,May_F3GT_2016}. Such mismatch may arise due to the presence of Fe-vacancies or defects in the real sample, which can alter the exchange interactions and hence the $T_\mathrm{C}$\cite{Parkin_2022}. For example, studies on bulk Fe$_{3}$GeTe$_{2}$ show $T_\mathrm{C}$ can vary from 220 K to 160 K depending on the concentration of Fe (as well as Ge and Te) atoms present in the sample\cite{Geetha_2021,May_F3GT_2016}


Therefore, one may conclude that the standard DFT technique gives reasonable results for magnetic moment and MAE, but overestimates the exchange interactions, especially the isotropic symmetric exchange, hence, leads to high values of the Curie temperature $T_\mathrm{C}$. While the magnetic moments, magnetic anisotropy energy and Curie temperature computed using DMFT are in a good agreement with the experimental studies reported in literature. The outcome from DFT+U calculations, for example, the magnetic moment, MAE and $T_\mathrm{C}$ are not at all compatible with experimental results. Therefore, the consideration of static electron correlation (in the limit of $U/W \gg 1$) can not be a reasonable choice for the correct description of the electronic and magnetic properties of FGT systems.



\subsection{Interatomic exchange interactions}\label{sec24}

Based on our results discussed in section~\ref{sec23}, we find that DMFT produces a reasonable outcome in terms of the magnetic properties of FGT monolayers. In this section, we discuss the isotropic symmetric and antisymmetric exchange interactions computed with the inclusion of dynamical electronic correlation effects. We focus on the exchange interactions obtained using DMFT, because, as discussed in section~\ref{sec23}, this technique captures the physical (electronic and magnetic) properties of FGT systems more accurately than other methods.

\subsubsection{Isotropic symmetric exchange parameters}\label{sec241}

     The isotropic symmetric exchange interactions $J_{ij}$ within the DFT+DMFT framework is given by\cite{Yaroslav_2015}:
    \begin{equation}
        J_{ij}=\frac{T}{4} \sum_{n} \mathrm{Tr}[\Delta_{i}(i\omega_{n})G_{ij}^{\uparrow}(i\omega_{n})\Delta_{j}(i\omega_{n})G_{ji}^{\downarrow}(i\omega_{n})],
        \label{Jij}
    \end{equation}
     where the trace is over the orbital degrees of freedom, $T$ is the temperature, and $\omega_{n}$ is the $n$th Matsubara frequency, $G_{ij}$ is the intersite Green's function between sites $i$ and $j$. Therefore, according to eq.~\ref{Jij}, the exchange interactions $J_{ij}$ depend on the onsite exchange splitting $\Delta_{i}$ and intersite Green's function $G_{ij}$.
     
     The onsite exchange splitting term $\Delta_{i}$ which includes the self-energy is given by:
        \begin{equation}
            \Delta_{i}(i\omega_{n})=H_{KS}^{\uparrow}+\Sigma_{i}^{\uparrow}(i\omega_{n})-H_{KS}^{\downarrow} - \Sigma_{i}^{\downarrow}(i\omega_{n}),
        \end{equation}
    where, $H_\mathrm{KS}$ and $\Sigma_{i}$ are the Kohn-Sham Hamiltonian and site-dependent self-energy. The self-energy is obtained by solving the DMFT equations. In DMFT calculations, the self-energy is frequency-dependent, which is also true for the exchange splitting. Though the self-energy $\Sigma$ is a single-site quantity, it affects the intersite Green’s function according to eq.~\ref{G} \cite{Yaroslav_2015}. $J_{ij}$ parameters for each Fe sublattice in Fe$_{n}$GeTe$_{2}$ ($n=3, 4, 5$) systems obtained using DMFT are plotted in Fig.~\ref{Jij_DMFT}.

    Figs.~\ref{Jij_DMFT} (a) and (b) show the $J_{ij}$ interactions when $i$=Fe1 and Fe3 for Fe$_{3}$GeTe$_{2}$ monolayer as a function of the nearest neighbor distance $R_{ij}$ (normalized by the lattice parameter $a$) obtained using DFT+DMFT. The dominating exchange interaction is ferromagnetic (FM) between different types of Fe species, i.e., when $i\neq j$ while for $i=j$, the first NN interaction is antiferromagnetic (AFM) irrespective of the type of Fe sublattice. As already mentioned in section~\ref{sec21}, the first NN distance between Fe1 and Fe2 is smaller than the first NN Fe1-Fe3 distance, therefore, as expected, the strongest FM $J_{ij}$ interaction takes place between the first NN Fe1-Fe2 pair, see Figs.~\ref{Jij_DMFT}(a) and (b). From the orbital decomposition of $J_{ij}$ interactions (see Table ~S1 in SI), we find the dominating contribution to the first nearest neighbor (NN) $J_{12}$ interaction comes from the $d_{{z}^{2}}$ orbital of the vertically aligned Fe1 and Fe2 atoms, see Fig.~\ref{fig1}(a).

    The $J_{ij}$ values calculated using standard DFT, DFT+DMFT and DFT+U techniques for Fe$_{3}$GeTe$_{2}$ monolayer are plotted in Fig.S3 SI. This comparison clearly shows that, although the sign of $J_{ij}$ interactions remains unchanged, their magnitude differs drastically, which in turn modifies the $T_\mathrm{C}$. We check the robustness of DFT+DMFT method with respect to the correlation strength by calculating the $J_{ij}$ interactions for different $U_\mathrm{eff} = U - J$ values using DFT+DMFT, for example, $U=5.0$ eV and $J_{H}=1.2$ eV, Seo et. al. considered these values in their studies on Fe$_{n}$GeTe$_{2}$ systems \cite{Sci_adv_FGT_family}. Our results show the maximum deviation in the largest $J_{ij}$ value is $\sim$8\% between the site-dependent $U_\mathrm{eff}$ values (calculated from cLR method) and $U=5.0, J_\mathrm{H}=1.2$ eV, for Fe$_{3}$GeTe$_{2}$ monolayer, see Fig.S4 SI.


    The $J_{ij}$ interactions obtained using DFT+DMFT for $i=$ Fe1 and Fe2 of Fe$_{4}$GeTe$_{2}$ monolayer are plotted in Figs.~\ref{Jij_DMFT}(c) and (d), respectively. Fig.~S5 shows the isotropic exchange interactions of Fe$_{4}$GeTe$_{2}$ computed using DFT, DFT+DMFT and DFT+U. In the case of Fe$_{4}$GeTe$_{2}$ as well, the sign of $J_{ij}$ couplings remain unchanged irrespective of the type of computational technique. The $J_{ij}$ interactions are mostly FM in this system and dominating FM interaction takes place between Fe1-Fe2 and Fe3-Fe4.

    The nature of $J_{ij}$ interactions is relatively complicated in the case of Fe$_{5}$GeTe$_{2}$. The dominating exchange interaction between different Fe sublattices is FM for $i=$ 2, 3, 4 and 5, while a small AFM coupling exists between the same Fe species, see Figs.~\ref{Jij_DMFT}(g)-(j). This scenario is quite different for $i=$1U and 1D, where significant AFM interactions are present even for $i \neq j$, i.e., between 1U or 1D and other Fe species, see Figs.~\ref{Jij_DMFT}(e) and (f). The presence of such non-negligible AFM coupling implies the existence of exchange frustration in Fe$_{5}$GeTe$_{2}$. Fig.~S6 shows how the absence and presence of electron correlation affect the $J_{ij}$ interactions present in Fe$_{5}$GeTe$_{2}$ for the UDU configuration. The $J_{ij}$ interactions computed using DFT and DFT+U in the UDU configuration are plotted in Fig.~S7 in SI.

     Using the DMFT computed $J_{ij}$ values we calculate exchange stiffness constant $A$ for the Fe$_{n}$GeTe$_{2}$ monolayers, which is given by:
    
    \begin{equation}
        A=\frac{1}{2V}\lim\limits_{\kappa\to0}\sum_{i \neq j} J_{ij}R_{ij}^{2}e^{-\kappa R_{ij}}
        \label{S1}
    \end{equation}
    where $V$ is the unit cell volume per magnetic site and $\kappa$ is the damping parameter. The term $e^{-\kappa R_{ij}}$ is introduced to improve the convergence of $A$ with the cutoff radii or $R_{ij}$\cite{Vladi_PRM_2022}. We check the dependence of $A$ as a function of $\kappa$ for different $R_{ij}$ varying $\kappa$ from 0 to 2 and we find for $\kappa \geq 0.5$ different curves coincide. Using those results we fit the exponential function $f(\kappa)=ae^{-b\kappa}+c$ and calculated $f(0)$ at the $\kappa = 0$ limit. The stiffness constant $A$ is estimated this way.

    Since we are interested in monolayer FGT systems, eq.~\ref{A} is modified in the following way:
    \begin{equation}
        A=\frac{1}{2S}\lim\limits_{\kappa\to0}\sum_{i \neq j} J_{ij}R_{ij}^{2}e^{-\kappa R_{ij}}
        \label{A}
    \end{equation}
    where $S$ is the unit cell area.

 The exchange stiffness constants $A$ for Fe$_{n}$GeTe$_{2}$ monolayers calculated using eq.~\ref{A} are given by: 4.63 meV ($n=3$), 4.30 meV ($n=4$), 3.24 meV ($n=5$, UUU) and 13.88 meV ($n=5$, UDU).

\subsubsection{Antisymmetric exchange interactions}\label{sec242}  
 
\begin{figure*} [t]
\begin{center}
\includegraphics[width=0.45\textwidth] {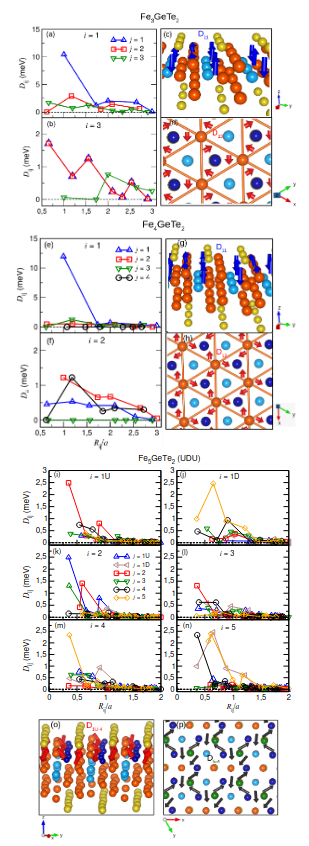}
\caption{\label{Dij_DMFT}Antisymmetric exchange parameters $D_{ij}=\sqrt{D_{x_{ij}}^{2}+D_{y_{ij}}^{2}+D_{z_{ij}}^{2}}$ as a function of neighbouring distance $R_{ij}$ for (a)-(b) Fe$_{3}$GeTe$_{2}$,  (e)-(f) Fe$_{4}$GeTe$_{2}$ and (i)-(n) Fe$_{5}$GeTe$_{2}$ monolayer, obtained using  DMFT. (c), (d), (g), (h), (o) and (p) show the first neighbor DM interactions between different $i$th and $j$th pairs. The $D_{ij}$ values are multiplied by the corresponding coordination numbers.}
\end{center}
\end{figure*}

Next, we study the antisymmetric exchange ($D_{ij}$) interactions. The $D_{ij}$ interactions are calculated using the relativistic generalization of the Lichtenstein-Katsnelson-Antropov-Gubanov (LKAG) formula\cite{LIECHTENSTEIN198765}. In the fully relativistic limit, the magnetic exchange parameters are ($3\times3$) tensors, with the isotropic exchange terms $J_{ij}$ as the diagonal components and the antisymmetric DM and the symmetric anisotropic exchange interactions in the off-diagonal components. 
     %
               
          

    Fig.~\ref{Dij_DMFT} shows the antisymmetric exchange interactions or Dzyaloshinkii-Moriya interaction (DMI) $D_{ij}=\sqrt{(D_{ij}^{x})^{2}+(D_{ij}^{y})^{2}+(D_{ij}^{z})^{2}}$ between different $i$ and $j$ in Fe$_{n}$GeTe$_{2}$ monolayers, computed using DMFT, where, $D_{ij}^{z}=\frac{1}{2}(J_{ij}^{xy}-J_{ij}^{yx})$.  The DM interactions for the two different Fe sublattices (Fe1 and Fe3) of Fe$_{3}$GeTe$_{2}$ are plotted in Figs.~\ref{Dij_DMFT}(a) and (b). The strongest $D_{ij}$ interaction takes place between Fe1-Fe1, i.e., Fe atoms situated in the same $z$ plane, as we see in Fig.~\ref{Dij_DMFT}(a) the first nearest neighbor $D_{11}$ interaction is $\sim$ 10 meV. Fig.~\ref{Dij_DMFT}(c) shows the side view of the first nearest neighbor $D_{11}$ acting between Fe1-Fe1 and the dominating contribution comes from the $D_{z}$ component. The next significant DM interaction takes place between the first nearest neighbors Fe1 and Fe3. Fig.~\ref{Dij_DMFT}(d) shows the top view of the DM interaction vectors between Fe1 and Fe3 sites, which have mainly the in-plane components ($D_{x}$ and $D_{y}$). It is important to note that the $D_{11}^{z}$ as well as the $D_{13}^{x}$ and $D_{13}^{y}$ vectors present in the unit cell almost cancel each other due to the structural symmetry of this system. More interestingly, the first nearest neighbour (NN) $D_{13}$ interaction is essentially zero, this happens because the NN Fe1 and Fe2 atoms are inversion symmetric partners with respect to the $z$ axis, see Fig.~\ref{fig1}(a). Comparing between the magnitudes of the highest $D_{ij}$ and $J_{ij}$ values of Fe$_{3}$GeTe$_{2}$ monolayer, we see the DM interaction is $\sim$3 times smaller than the isotropic symmetric exchange interactions. Fig.~S9 in SI shows the DM interactions present in Fe$_{3}$GeTe$_{2}$ obtained using DFT and DFT+DMFT.

    The $D_{ij}$ interactions for different Fe sublattices of Fe$_{4}$GeTe$_{2}$ are plotted in Fig.~\ref{Dij_DMFT}(e) and (f). Similar to Fe$_{3}$GeTe$_{2}$, in this case also the first NN $D_{11}$ interaction (between Fe1 and Fe1) has the highest contribution to DMI in Fe$_{4}$GeTe$_{2}$ with $D_{z}$ as the dominating component, see Fig.~\ref{Dij_DMFT}(g). The next dominating DM interaction takes place between the first NN Fe2-Fe2 pair, and the direction of $D_{22}$ is mainly along $z$. There exists non-zero DMI between Fe1-Fe2 or Fe3-Fe4 where the in-plane components dominate, see Fig.~\ref{Dij_DMFT}(h). The first nearest neighbor $D_{13}$ or $D_{24}$ interaction is zero, while the long-range $D_{13}$ (or $D_{24}$) are non-zero. More interestingly, DM interactions do not exist between Fe1 and Fe4  (Fe2 and Fe3), these are the inversion symmetric partners, see Fig.~\ref{fig1}(b). The DM interaction vectors between different Fe-sublattices of Fe$_{4}$GeTe$_{2}$ are plotted in Figs.~\ref{Dij_DMFT}(g) and (h).  Fig.~S10 shows the DM interactions present in Fe$_{4}$GeTe$_{2}$ obtained using DFT and DFT+DMFT.

    Figs.~\ref{Dij_DMFT}(i)-(n) show the DM interactions for different Fe sublattices of Fe$_{5}$GeTe$_{2}$ in UDU configuration. The $D_{ij}$ interactions are quite complex in this case compared to the Fe$_{3}$GeTe$_{2}$ and Fe$_{4}$GeTe$_{2}$ monolayers, because this system is less symmetric than the first two FGT systems. Figs.~\ref{Dij_DMFT}(o) and (p) show the DM vectors between Fe1U-Fe4 and Fe4-Fe5 pairs in Fe$_{5}$GeTe$_{2}$.  Fig.~S11 shows the DM interactions present in Fe$_{5}$GeTe$_{2}$ monolayer in UUU configuration computed using DFT and DFT+DMFT.
  
  Though the scalar form of $D_{ij}$ interactions plotted in Fig.~\ref{Dij_DMFT} shows a non-zero contribution for some of the site neighbors, the net DMI for all the nearest neighbors of a given site $i$ is quite small due to the structural symmetries present in the pristine FGT monolayers. Using the computed $\vec{D}_{ij}$ values we calculate the spiralization constant $D$ for the Fe$_{n}$GeTe$_{2}$ monolayers defined as:
    
    \begin{equation}
        D=\frac{1}{S}\lim\limits_{\kappa\to0}\sum_{i \neq j} \vec{D}_{ij}\otimes \vec{R}_{ij}e^{-\kappa R_{ij}}
        \label{D}
    \end{equation}
    In this case $D$ is $3\times3$ tensor. Since our system of interest is 2D FGT systems, we divide eq.~\ref{D} by the unit cell area $S$ (same as eq.~\ref{A}).

   The spiralization constants calculated from the DM vectors are negligible in the case of Fe$_{3}$GeTe$_{2}$ and Fe$_{4}$GeTe$_{2}$, because the net DM vector acting between any sites $i$ and $j$ gets significantly cancelled out due to the structural symmetries, as we see in Figs.~\ref{Dij_DMFT}(c), (d), (g) and (h).
   In the case of Fe$_{5}$GeTe$_{2}$ in UUU and UDU configurations, the diagonal terms in eq.~\ref{D} are very small but there are some non-negligible off-diagonal terms. The highest off-diagonal term of the spiralization tensor for UUU and UDU configurations are
    0.04 meV \AA$^{-1}$ and 0.031 meV \AA$^{-1}$, respectively. Due to the existence of such small DM interactions in Fe$_{n}$GeTe$_{2}$ monolayers, there is a negligible contribution to $T_\mathrm{C}$ from the antisymmetric exchange interactions.




\subsection{Electronic structure}\label{sec25}


    In order to investigate how electron correlation effect modifies the electronic structure we compare the DFT band structure and $k$-resolved spectral density obtained from DMFT, along the high symmetry directions. We also discuss the difference in density of states for these two cases.

   The DMFT expressions for spectral density $A(\textbf{k},\epsilon)$ and density of states (DOS) $D(\epsilon)$ are given by: 
    
     \begin{equation}
         A(\textbf{k},\epsilon)=-\frac{1}{\pi}\sum_{\chi}<k,\chi|\mathrm{Im}G(\epsilon+i0)|k,\chi>,
      \end{equation}
      
       \begin{equation}
        D(\epsilon)=-\frac{1}{\pi}\mathrm{Tr}[\mathrm{Im} G(\epsilon+i0)],
        \label{DOS}
   \end{equation}
       
\begin{figure*} [t]
\begin{center}
\includegraphics[width=1.01\textwidth] {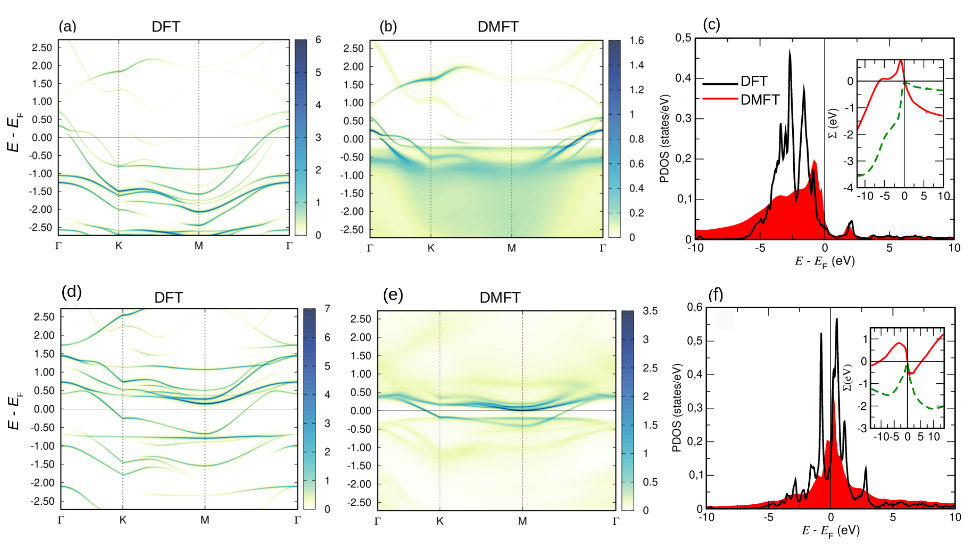}
\caption{\label{BS_DOS} Band structure projecting on both spin channels of $d_{z^{2}}$ orbital for Fe1 (Fe2) in Fe$_{3}$GeTe$_{2}$ using (a), (d) DFT and (b), (e) DMFT. (c) and (f) show the DFT (black) vs. DFT+DMFT (red) density of states projected on $d_{z^{2}}$ orbital of Fe1(Fe2) for both spin channels. Insets in (c) and (f) show the real (red solid) and imaginary (green dashed) parts of the self-energy $\Sigma$ for both spin channels. }
\end{center}
\end{figure*}

    Among the three Fe$_{n}$GeTe$_{2}$ systems, $n=3$ is the simplest one and the $J_{ij}$ interactions present in this system are less complicated. Hence, we focus only on Fe$_{3}$GeTe$_{2}$, to see the change in electronic structure due to dynamical electronic correlations. Here we focus on Fe1-$d_{{z}^{2}}$ state since from the orbital resolved analysis of $J_{ij}$ interactions we see this particular orbital of Fe1 (and/or Fe2) has dominating contribution to the isotropic symmetric exchange interaction $J_{12}$ (between Fe1 and Fe2), as reported in section~\ref{sec241} and Table~S1 in SI. Fig.~\ref{BS_DOS} shows the comparison of electronic structures of Fe$_{3}$GeTe$_{2}$ monolayer obtained without and with dynamical correlations.

    The spin-polarized DFT and DFT+DMFT band structures for Fe1 (or Fe2) sublattice of Fe$_{3}$GeTe$_{2}$ projecting on $d_{{z}^{2}}$ orbital are plotted along the high-symmetry directions in Fig.~\ref{BS_DOS}(a)-(b) (up spin) and (d)-(e) (down spin), respectively. The DFT band structure contains sharp energy bands, while the DMFT bands are smeared out due to the finite quasiparticle lifetime\cite{Wills2000} induced by a non-zero imaginary part of the self-energy. Furthermore, the energy levels are shifted towards the Fermi level, which is attributed to the real part of the self-energy, and become more flat due to correlations. Correlations also reduce the bandwidth in the momentum-resolved spectral function $A(k, \epsilon)$. In Fig.~\ref{Jij_DMFT}(a) we see the first NN $J_{12}$, i.e., the isotropic symmetric exchange interaction between Fe1 and Fe2, is the strongest among all $J_{ij}$ interactions present in this system. The orbital decomposed results show the first NN Fe1 and Fe2 atoms interact mainly via their $d_{{z}^{2}}$ orbitals. Due to this reason, here we focus on the $d_{{z}^{2}}$ orbital of one of these Fe species, and compare its electronic nature using DFT and DFT+DMFT. Comparing between the DFT and DMFT band structures we find the narrowing or shift of bands towards the Fermi energy is quite significant along $\Gamma-K$ and $\Gamma-M$ directions for the spin-up and down channels, respectively.
    

    The DFT (black) and DFT+DMFT (red) density of states $D(\epsilon)$ projecting on Fe1 (or Fe2) $d_{{z}^{2}}$ orbital for both spin channels are plotted in Figs.~\ref{BS_DOS}(c) and (f). It follows from Figs.~\ref{BS_DOS}(c) and (f) that DMFT causes noticeable renormalization of Fe1-3$d_{{z}^{2}}$ projected density of states (PDOS). In the case of spin-up channel of the DMFT PDOS (red curve), the first peak is situated closer to the Fermi energy compared to the DFT PDOS, see Fig.~\ref{BS_DOS}(c). Similarly, for the spin-down channel the electronic states are also shifted toward the Fermi level for DMFT PDOS, see Fig.~\ref{BS_DOS}(f). 

    Even though the integration of spin-polarized DOS up to the Fermi energy remains almost the same for both DFT and DMFT, (which causes similar value of magnetic moments for DFT and DMFT), the electronic spectra gets modified upon the inclusion of dynamical electronic correlations. These modifications in the PDOS can be understood by analyzing the real (red solid) and imaginary (green dashed) parts of the self-energy $\Sigma$, displayed as the insets in Figs.~\ref{BS_DOS}(c) and (f), for the up and down spin channels, respectively. According to eq.~\ref{DOS} the DMFT DOS contains imaginary part of the Green's function, and the self energy term $\Sigma$ is included in the denominator of Green's function, see eq.~\ref{G}. The quasiparticle energies are renormalized by the real part of the self-energy which shifts the positions of the peaks in the DMFT spectrum. The high and positive value of the self-energy for the up and down spin channels within the energy range between 0 and -5 eV shifts the spectrum toward the Fermi level. 
    The imaginary part of the self-energy causes broadening of the peaks present in PDOS, leading to an effective decrease of their intensities. Note, the imaginary part of the self energy is $\sim$0 at and close to the Fermi energy, implying the quasi-particles have longer lifetimes, according to Fermi liquid theory.

    The difference in electronic structure between DFT and DMFT, displayed in Fig.~\ref{BS_DOS}, is responsible to modify the isotropic symmetric exchange interactions present in Fe$_{3}$GeTe$_{2}$ monolayer, which in turn changes the Curie temperature. The band structure plot along high-symmetry directions shows that the energy levels get more localized in the case of DMFT than DFT, thus the dynamical correlations reduce the electron hopping amplitude between different sites and consequently the intersite Green's-function $G_{ij}$ in eq.~\ref{Jij}. Therefore, the isotropic symmetric exchange interactions present in FGT systems get significantly reduced in DMFT leading to smaller values of $T_\mathrm{C}$.

    


\subsection{Discussion on Bulk FGT systems \label{sec25}}


    $T_\mathrm{C}$ obtained from our Monte Carlo simulations for Fe$_{3}$GeTe$_{2}$, Fe$_{4}$GeTe$_{2}$ and Fe$_{5}$GeTe$_{2}$ (UDU configuration) monolayers are 275 K, 220 K and 280 K, respectively. From these results one can see that $T_\mathrm{C}$ does not increase linearly as a function of $n(=3, 4,$ and 5), which is in contrast to previous studies on bulk Fe$_{n}$GeTe$_{2}$ systems\cite{Sci_adv_FGT_family,Ramesh_PRB_2020}. This implies some important features must be missing in the case of FGT monolayers compared to the bulk FGT systems. Due to the lack of those features, the expected $T_\mathrm{C}$ vs. $n$ relationship is not observed.

    To understand the real scenario present in bulk FGT systems we focus on bulk Fe$_{3}$GeTe$_{2}$ since this is the simplest member. Based on our (DFT and DFT+DMFT) results we find (as mentioned in section~\ref{sec23}) that the value of $T_\mathrm{C}$ primarily depends on $J_{ij}$. Therefore, the sign and magnitude of $J_{ij}$ interactions present in bulk Fe$_{3}$GeTe$_{2}$ should be the main features to tune the $T_\mathrm{C}$ from monolayer to bulk (or vice versa). We calculate the magnetic exchange interactions and $T_\mathrm{C}$ for bulk Fe$_{3}$GeTe$_{2}$. Our results show both the strength and magnitude of intralayer $J_{ij}$ interactions remain almost the same for monolayer and bulk, see Fig. S12 in SI. Therefore the intralayer $J_{ij}$ interactions do not play any role in modifying the $T_\mathrm{C}$ for this system. 

\begin{figure*} [t]
\centering
\includegraphics[width=1.01\textwidth] {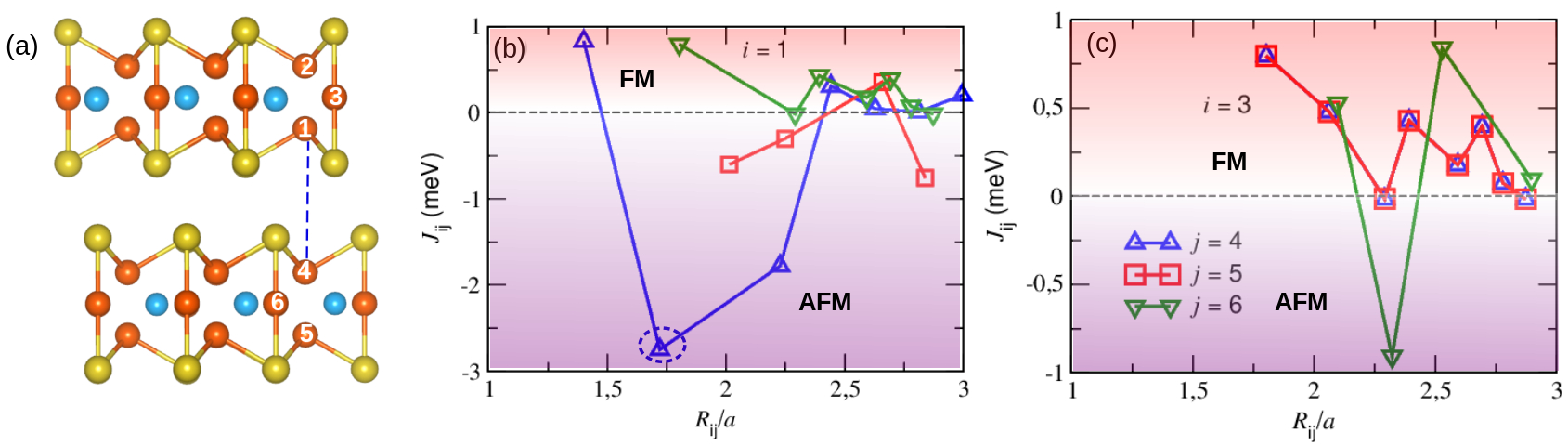}
\caption{(a) Side view of bulk Fe$_{3}$GeTe$_{2}$, the sublatiices highlighted in white show the strongest AFM interaction. Interlayer symmetric exchange couplings $J_{ij}$ as a function of neighbouring distance $R_{ij}$ for Fe$_{3}$GeTe$_{2}$ bulk for the $i$th site as (b) Fe1 and (c) Fe3. These results are obtained using DFT+DMFT. The second nearest neighbor $J_{14}$ (or $J_{36}$) interaction shows significant AFM coupling. The exchange parameters are multiplied by the corresponding coordination numbers. The blue dashed line in (a) between Fe1 and Fe4 shows the Fe sublattices taking part into the highest AFM interaction, highlighted by the blue circle in (b). }
\label{Bulk_F3GT}
\end{figure*}

   Next, we focus on the interlayer symmetric exchange interactions and our DFT+DMFT results show that significant antiferromagnetic (AFM) couplings exist in bulk Fe$_{3}$GeTe$_{2}$, as already discussed in the literature\cite{Yi_2016,Kim_2019,Nanoscale_2020}.
   Fig.~\ref{Bulk_F3GT}(a) and (b), (c) show the side view of bulk Fe$_{3}$GeTe$_{2}$ and the interlayer symmetric exchange interactions $J_{ij}$ for $i=1$ and $i=3$, respectively. Bulk Fe$_{3}$GeTe$_{2}$ has the vdW gap of 3.47 \AA\ and the out-of-plane lattice constant is 16.33 \AA. Note, the unit cell of bulk Fe$_{4}$GeTe$_{2}$ and Fe$_{5}$GeTe$_{2}$ contain three formula units along $z$ direction with thickness 29.08 and 29.20 \AA, respectively\cite{Sci_adv_FGT_family,may2019ferromagnetism}. Fig.~\ref{Bulk_F3GT}(b) and (c) significant interlayer AFM interactions are present in bulk Fe$_{3}$GeTe$_{2}$. Though these interlayer $J_{ij}$ interactions are weaker than the intralayer interactions, such exchange couplings are enough to tune the $T_\mathrm{C}$. As a result, the Curie temperature changes from 260 K (monolayer) to 205 K (bulk), and this value of $T_\mathrm{C}$ for the pristine bulk Fe$_{3}$GeTe$_{2}$ is in excellent agreement with the previously reported experimental findings\cite{Geetha_2021,Sciadv_F3GT_2018,Sci_adv_FGT_family,May_F3GT_2016,Inorg_F3GT_2015,npj_F3GT} and computed values of $T_\mathrm{C}$\cite{ZHU2021,Sci_adv_FGT_family}.
   Note, from standard DFT calculations the value of $T_\mathrm{C}$ for bulk Fe$_{3}$GeTe$_{2}$ is found to be 410 K, which is hugely overestimated compared to DFT+DMFT and also w.r.t. the experimental reports. Fig.S13 shows the comparison in magnetization ($M$) vs. temperature ($T$) behavior for bulk Fe$_{3}$GeTe$_{2}$ between DFT and DFT+DMFT.

    Similar to Fe$_{3}$GeTe$_{2}$, if we compute $T_\mathrm{C}$ of bulk Fe$_{4}$GeTe$_{2}$ and Fe$_{5}$GeTe$_{2}$ (UDU configuration), then we should in principle obtain the linear relationship between $T_\mathrm{C}$ and $n$ ($n=3, 4, 5$). Since we have already discussed in this study that DFT+DMFT describes the physical properties of FGT systems more accurately than standard DFT or DFT+U, therefore one has to perform DFT+DMFT calculations for bulk Fe$_{4}$GeTe$_{2}$ and Fe$_{5}$GeTe$_{2}$ to get compatible results with experiment. But these DMFT calculations are computationally expensive due to a large number of atoms. Therefore, in this study, we have limited ourselves to DFT+DMFT calculations only for bulk  Fe$_{3}$GeTe$_{2}$. Based on our results and physical interpretation we expect that the interlayer exchange interactions present in other bulk FGT systems would modify the corresponding $T_\mathrm{C}$ w.r.t. the monolayer and one can probably expect a linear trend for $T_\mathrm{C}$ vs. $n$ similar to experiment.


\section{Conclusion}\label{sec3}
In this study, we have performed a systematic investigation of the electronic structure and magnetic properties of Fe$_{n}$GeTe$_{4}$ ($n=3, 4,$ and 5) systems by DFT, DFT+U and DMFT methods. By using the computed parameters in the spin Hamiltonian, the Curie temperature for each FGT monolayer was determined. Based on our results, it is quite evident that, among these three methodologies, DFT+U with static electronic correlations is not applicable for these systems, since it produces values of magnetic moments and Curie temperature that do not agree with the experiment. We find that the magnetic moments obtained using standard DFT are compatible with the experimental results, but the Curie temperature is overestimated. Upon the inclusion of dynamical electronic correlations within the DFT+DMFT approach, the magnetic moments remain almost unchanged, while the exchange interactions, especially the isotropic symmetric exchange parameters get significantly modified to decrease the Curie temperature substantially. The latter is in a good agreement with experimental reports for all FGT systems. This implies that consideration of dynamical correlations is necessary to capture the correct electronic structure and magnetic behavior of the FGT systems.

\section{Methods and Computational Details}\label{sec4}
Density functional theory (DFT) based calculations are performed to study the structural, electronic, and magnetic properties of Fe$_{n}$GeTe$_{2}$ (FGT family) systems. Structural optimizations are performed using Vienna Ab initio Simulation Package (VASP)\cite{Kresse1999, Kresse1994}, where the exchange-correlation potential has been treated with the generalized gradient approximation (GGA), with the Perdew-Burke-Ernzerhof (PBE) functional\cite{Perdew1996}. We use 18$\times$18$\times$1 Monkhorst-Pack $k$-point mesh in our calculations for Brillouin zone (BZ) integration.\cite{Monkhorst1976} To model isolated 2D monolayers, the interaction between periodic images of the supercell along the $z$-axis was reduced by adding a 20\,{\AA} vacuum region perpendicular to the surface of monolayers. The lattice constants and atomic coordinates were optimized by minimizing energy based on the conjugate gradient method with a force component tolerance of 0.01 eV/\AA\ on each atom. The energy cutoff for the plane-wave basis set was set to 500 eV.

To have more accurate prediction of localized magnetic properties, we used the full-potential linear muffin-tin orbital (FP-LMTO) method, implemented in the RSPt code\cite{Wills1987,Wills2000}. The $k$-point grids with dimensions 21$\times$21$\times$1 and 36$\times$36$\times$1 were constructed for calculating the magnetic exchange interactions ($J_{ij}$) and magnetic anisotropy energy (MAE), respectively.
Within the scope of FP-LMTO, we studied the electronic and magnetic properties of FGT systems using different computational approaches: DFT, dynamical mean-field theory (DMFT) and DFT+U.


\subsection{DMFT methodology}\label{sec41}

The key concept of DMFT is that the Hubbard model is mapped locally to an effective Anderson impurity model, where the problem of an entire lattice is converted into the simple problem of an atom embedded in an electronic bath. Each lattice-site is coupled with the bath which represents rest of the crystal and electron on the single
site could be created or annihilated by coupling with the electronic bath\cite{Georges_DMFT}. The Hamiltonian of the Anderson impurity model can be written as:

\begin{equation}
H_{AIM}=H_{atom}+\sum_{\nu,\sigma}\epsilon_{\nu}a^{\dagger}_{\nu\sigma}a_{\nu\sigma}+\sum_{\nu\sigma}(V_{\nu}c^{\dagger}_{0\sigma}a_{\nu\sigma}+h.c.)+Un_{0\uparrow}n_{0\downarrow},
 \label{H_DMFT}
\end{equation}

 where $H_\mathrm{atom}$ includes the single site interactions and $\epsilon_{\nu}$ is the energy associated with the bath electrons. The first two terms can be considered as the non-interaction energy terms. While the third term describes the coupling between the bath electrons and the electron on the single atom site, where $V_{\nu}$ is the strength of this coupling. ($c^{\dagger}_{0,\sigma},c_{0,\sigma}$) and ($a^{\dagger}_{\nu,\sigma},a_{\nu,\sigma}$) are the degrees of freedom for the on-site and bath electrons, respectively. The last term describes the local Coulomb interaction $U$.



The converged DFT calculations are the starting point to perform our DMFT calculations as implemented in RSPt\cite{Wills2000,Oscar}. The first step of the DFT+DMFT or DMFT method is to identify a set of local orbitals, which are not properly described by the standard DFT technique. The Hamiltonian written in eq.~\ref{H_DMFT} can be solved through DMFT, and convergence should be achieved for both the local self-energy and the full electron density. To solve the effective impurity problem arising in the DMFT cycle we use the spin polarized T-matrix fluctuation-exchange (SPTF) solver\cite{Wills2000}. SPTF solver is chosen for our calculations due to its efficiency and accuracy for the moderately correlated systems\cite{Eriksson_PRB_2007,Lichenstein_2005}. The double counting correction $H_{DC}$ is considered as the orbitally averaged static part of the self-energy, which is usually done for the SPTF solver\cite{Kotilar_PRL_2001,Eriksson_PRB_2007}. In our calculations we use full-potential muffin-tin orbitals as implemented in the RSPt code. The Fermi smearing is used for the electronic occupations for the FGT systems, with the smearing temperature 155 K. The effective Hubbard parameter $U_\mathrm{eff}$ used in our calculations are obtained from the constrained linear response method, as described in section~\ref{sec23}.

\subsection{Methodology for $J_{ij}$ calculations}\label{sec42}

The isotropic symmetric exchange interactions $J_{ij}$ are calculated within the full-potential linearized muffin-tin orbital (FPLMTO) basis implemented in the RSPt code. From the LMTO basis, one can construct the Bloch sums to solve the DFT eigenvalue problem and subsequently for the one-electron Green’s function. We have considered L{\"o}wdin orthonormalized LMTO basis functions, which are not very localized due to their long decaying tail, and are more physical for metallic systems like Fe$_{5}$GeTe$_{2}$. For a detailed description of the shape of various local orbitals used in RSPt, please see Ref.~\citenum{Yaroslav_2015}. The L{\"o}wdin orbitals used in our calculations have been constructed from the original LMTO basis functions performing a k-point-wise orthonormalization\cite{Yaroslav_2015}. 

$J_{ij}$ can be extracted from the Green's function obtained from the LMTO basis. The generalized expression for the intersite exchange parameters is given by:

\begin{equation}
    J_{ij} = \frac{T}{4}\sum_{n} [\hat{\Delta}_{i} (i\omega_{n})\hat{G}_{ij}^{\uparrow}\hat{\Delta}_{j} (i\omega_{n})\hat{G}_{ji}^{\downarrow} ],
    \label{Jij_DFT}
\end{equation}

The most important quantities in this expression are the onsite spin splitting $\Delta_{i}$ and the spin-dependent intersite Green's function $\hat{G}_{ij}$. The trace in eq.~\ref{Jij_DFT} is taken over the orbital degrees of freedom. $T$ and $\omega_{n}=2\pi T(2n+1)$ are the temperature and the $n$th fermionic Matsubara frequency respectively. $\hat{G}_{ij}^{\sigma}$ is the intersite Green’s function between sites $i$ and $j$ and projected over a given spin $\sigma$. The term $\Delta_{i}$ gives the exchange splitting at site $i$, obtained using spin and site-projected Kohn-Sham Hamiltonian.
In the presence of dynamical electronic correlations, eq.~\ref{Jij_DFT} gets modified and in DMFT the $J_{ij}$ parameters are calculated using eq.~\ref{Jij} as mentioned in section~\ref{sec241}.

\subsection{Methodology for $D_{ij}$ calculations}\label{sec43}
In order to obtain information on more complex magnetic interactions, one has to generalize this approach to the relativistic case, which has been done already
within the Korringa-Kohn-Rostoker (KKR) Green-function method\cite{Kvashnin2020}. 
The details concerning the implementation of antisymmetric exchange interactions in RSPt and the calculation results for different correlated systems are discussed in refs.~\citenum{Kvashnin2020} and \citenum{Vladislav_PRB_2021}.

\subsection{Monte Carlo simulations}\label{sec44}

To estimate the magnetic ordering temperatures, we performed classical Monte Carlo (MC) simulations via UppASD code\cite{eriksson2017atomistic}, where the calculated magnetic parameters are implemented in the Hamiltonian introduced in eq.~\ref{H}. Here, identical $K_{i}$ was assumed for all Fe sites by averaging the total MAE/cell by the number of Fe atoms present in the unit cell. To achieve properly averaged properties, calculations were done for three ensembles in supercell with sizes varying between 40$\times$40$\times$1, 50$\times$50$\times$1, and 60$\times$60$\times$1, where periodic boundary conditions were imposed along $x$ and $y$ axes, and the transition temperatures were estimated by monitoring the cross sections of fourth-order cumulants of magnetization.

\subsection{Constrained Linear Response Method}\label{sec45}
We computed the Hubbard $U$ parameter by means of constrained-density-functional calculations. The Hubbard $U$ was computed by varying the electron occupation of a single site by constructing a supercell where the periodically repeated sites are perturbed coherently. In this supercell approach, the occupation of one representative site in a large cell is changed leaving all other site occupations unchanged.

The effective interaction parameter $U$ associated to site $i$ in terms of the response function $\chi$ is given by: $U=(\chi_{0}^{-1}-\chi^{-1})_{ii}$ 

We performed a well converged self-consistent field DFT calculation for the unperturbed system for all sites in the supercell. Then starting from its self-consistent potential, small positive and negative potential shifts were added to each nonequivalent “Hubbard” site $j$ and the variation of the occupations for all the $i$ sites present in the supercell was computed in two steps: i) re-adjusting the Kohn-Sham potential of the system self-consistently to optimally screen the localized perturbation, and ii) without allowing this screening. This latter result is nothing but the variation computed from the first iteration in the self-consistent cycle leading eventually to the former (screened) results. The site-occupation derivatives calculated according to i) and ii) produce the matrices $\chi_{ij}$ and $\chi_{ij}^{0}$, respectively. The difference between these matrices gives the value of the Hubbard parameter $U$.

This scheme is rotationally invariant, therefore the Hund's exchange $J_\mathrm{H}$, describing these effects can be considered as zero, or the effect of $J_\mathrm{H}$ can be mimicked by redefining the $U$ parameter as $U_\mathrm{eff} = U-$ $J$, which is known as Dudarev scheme in literature\cite{Dudarev}.


\textbf{Supplementary information}
Supporting Information accompanies this article which is available free of charge.



\textbf{Acknowledgments}
B.S. and S.G. acknowledge a postdoctoral grant from Carl Tryggers Stiftelse (CTS 20:378). The computations were enabled in project SNIC 2021/3-38 by resources provided by the Swedish National Infrastructure for Computing (SNIC) at NSC, PDC, and HPC2N partially funded by the Swedish Research Council (Grant No. 2018-05973). B.S. acknowledges allocation of supercomputing hours by PRACE DECI-17 project `Q2Dtopomat' in Eagle supercompter in Poland and EuroHPC resources in Karolina supercomputer in Czech Republic.

\textbf{Author contributions}
S.G. and S.E. carried out the calculations. All authors discussed the results and co-wrote the paper.

\textbf{Correspondence}
Correspondence to Biplab Sanyal.

\textbf{Competing interests}
The authors declare no competing interests.

\textbf{Data availability}
The data supporting the findings of this study are available from the corresponding authors upon request.

\textbf{Code availability}
The structural optimizations are performed using VASP. The full-potential LMTO code RSPt was used to compute the magnetic and electronic properties. The Monte Carlo simulations are performed using the UppASD code. All these packages or codes are commercially available.

\section{References}
\bibliography{sn-bibliography}

\providecommand{\newblock}{}
\begin{thebibliography}{10}
\expandafter\ifx\csname url\endcsname\relax
  \def\url#1{{\tt #1}}\fi
\expandafter\ifx\csname urlprefix\endcsname\relax\def\urlprefix{URL }\fi
\providecommand{\eprint}[2][]{\url{#2}}

\bibitem{Novoselov_2007}
Geim A~K and Novoselov K~S 2007 {\em Nat. Mater.\/} {\bf 6}(3) 183--191

\bibitem{Nat_rev_phys_2019}
Mak K~F, Shan J and Ralph D~C 2019 {\em Nat. Rev. Phys.\/} {\bf 1}(11) 646--661

\bibitem{Nat_fm_2017}
Gong C, Li L, Li Z, Ji H, Stern A, Xia Y, Cao T, Bao W, Wang C, Wang Y, Qiu
  Z~Q, Cava R~J, Louie S~G, Xia J and Zhang X 2017 {\em Nature\/} {\bf
  546}(7657) 265--269

\bibitem{mermin1966absence}
Mermin N~D and Wagner H 1966 {\em Phys. Rev. Lett.\/} {\bf 17} 1133

\bibitem{Nature_2020}
Huang B, Clark G, Navarro-Moratalla E, Klein D~R, Cheng R, Seyler K~L, Zhong D,
  Schmidgall E, McGuire M~A, Cobden D~H, Yao W, Xiao D, Jarillo-Herrero P and
  Xu X 2017 {\em Nature\/} {\bf 546}(7657) 270--273

\bibitem{CrI3_npj}
Xu C, Feng J, Xiang H and Bellaiche L 2018 {\em npj. Comput. Mater.\/} {\bf
  4}(1) 57

\bibitem{Adv_Mater_2020}
Cortie D~L, Causer G~L, Rule K~C, Fritzsche H, Kreuzpaintner W and Klose F 2020
  {\em Adv. Funct. Mater.\/} {\bf 30} 1901414

\bibitem{vzutic2004spintronics}
{\v{Z}}uti{\'c} I, Fabian J and Sarma S~D 2004 {\em Rev. Mod. Phys.\/} {\bf 76}
  323

\bibitem{cortie2020two}
Cortie D~L, Causer G~L, Rule K~C, Fritzsche H, Kreuzpaintner W and Klose F 2020
  {\em Adv. Funct. Mater\/} {\bf 30} 1901414

\bibitem{gong2017discovery}
Gong C, Li L, Li Z, Ji H, Stern A, Xia Y, Cao T, Bao W, Wang C, Wang Y {\em
  et~al.\/} 2017 {\em Nature\/} {\bf 546} 265--269

\bibitem{huang2017layer}
Huang B, Clark G, Navarro-Moratalla E, Klein D~R, Cheng R, Seyler K~L, Zhong D,
  Schmidgall E, McGuire M~A, Cobden D~H {\em et~al.\/} 2017 {\em Nature\/} {\bf
  546} 270--273

\bibitem{BHATTI2017530}
Bhatti S, Sbiaa R, Hirohata A, Ohno H, Fukami S and Piramanayagam S 2017 {\em
  Mater. Today\/} {\bf 20} 530--548 ISSN 1369-7021

\bibitem{Tsymbal_2019}
Li X, L{\"u} J~T, Zhang J, You L, Su Y and Tsymbal E~Y 2019 {\em Nano Lett.\/}
  {\bf 19}(8) 5133--5139

\bibitem{Tsymbal_2022}
Huang K, Shao D~F and Tsymbal E~Y 2022 {\em Nano Lett.\/} {\bf 22}(8)
  3349--3355

\bibitem{Zhang_Adv_Matr_2019}
Zhang Y, Chu J, Yin L, Shifa T~A, Cheng Z, Cheng R, Wang F, Wen Y, Zhan X, Wang
  Z and He J 2019 {\em Adv. Mater.\/} {\bf 31} 1900056

\bibitem{FeTe_2020}
Kang L, Ye C, Zhao X, Zhou X, Hu J, Li Q, Liu D, Das C~M, Yang J, Hu D, Chen J,
  Cao X, Zhang Y, Xu M, Di J, Tian D, Song P, Kutty G, Zeng Q, Fu Q, Deng Y,
  Zhou J, Ariando A, Miao F, Hong G, Huang Y, Pennycook S~J, Yong K~T, Ji W and
  Renshaw~Wang X 2020 {\em Nat. Commun.\/} {\bf 11} 3729

\bibitem{Abdullahi_2021}
Abdullahi Y~Z, Vatansever Z~D, Ersan F, Akinci U, Akturk O~U and Akturk E 2021
  {\em Phys. Chem. Chem. Phys.\/} {\bf 23}(10) 6107--6115

\bibitem{Wu_2021}
Wu H, Zhang W, Yang L, Wang J, Li J, Li L, Gao Y, Zhang L, Du J, Shu H and
  Chang H 2021 {\em Nat. Commun.\/} {\bf 12}(1) 5688

\bibitem{Meng_2021}
Meng L, Zhou Z, Xu M, Yang S, Si K, Liu L, Wang X, Jiang H, Li B, Qin P, Zhang
  P, Wang J, Liu Z, Tang P, Ye Y, Zhou W, Bao L, Gao H~J and Gong Y 2021 {\em
  Nat. Commun.\/} {\bf 12}(1) 809

\bibitem{Ramesh_PRB_2020}
Zhang H, Chen R, Zhai K, Chen X, Caretta L, Huang X, Chopdekar R~V, Cao J, Sun
  J, Yao J, Birgeneau R and Ramesh R 2020 {\em Phys. Rev. B\/} {\bf 102}(6)
  064417

\bibitem{may2019ferromagnetism}
May A~F, Ovchinnikov D, Zheng Q, Hermann R, Calder S, Huang B, Fei Z, Liu Y, Xu
  X and McGuire M~A 2019 {\em ACS nano\/} {\bf 13} 4436--4442

\bibitem{Ly_Adv_Mater_2021}
Ly T~T, Park J, Kim K, Ahn H~B, Lee N~J, Kim K, Park T~E, Duvjir G, Lam N~H,
  Jang K {\em et~al.\/} 2021 {\em Adv. Funct. Mater.\/} {\bf 31} 2009758

\bibitem{liu2022layer}
Liu Q, Xing J, Jiang Z, Guo Y, Jiang X, Qi Y and Zhao J 2022 {\em Commun.
  Phys.\/} {\bf 5} 1--10

\bibitem{JOE2019299}
Joe M, Yang U and Lee C 2019 {\em N. Mater. Sci.\/} {\bf 1} 299--303 ISSN
  2589-9651

\bibitem{F3GT_DFT_2016}
Zhuang H~L, Kent P~R~C and Hennig R~G 2016 {\em Phys. Rev. B\/} {\bf 93}(13)
  134407

\bibitem{Stoner}
Clifton S~E 1938 {\em Proc. R. Soc. Lond. Ser. A. Math. Phys. Sci.\/} {\bf 165}
  372--414

\bibitem{F3GT_JPCL}
Wang H, Xu R, Liu C, Wang L, Zhang Z, Su H, Wang S, Zhao Y, Liu Z, Yu D, Mei
  J~W, Zou X and Dai J~F 2020 {\em J. Phys. Chem. Lett.\/} {\bf 11}(17)
  7313--7319

\bibitem{Non_Stoner}
Xu X, Li Y~W, Duan S~R, Zhang S~L, Chen Y~J, Kang L, Liang A~J, Chen C, Xia W,
  Xu Y, Malinowski P, Xu X~D, Chu J~H, Li G, Guo Y~F, Liu Z~K, Yang L~X and
  Chen Y~L 2020 {\em Phys. Rev. B\/} {\bf 101}(20) 201104

\bibitem{Blugel_2022}
Schmitt M, Denneulin T, Kovács A, Saunderson T~G, R{\"u}$\beta$mann P, Shahee
  A, Scholz T, Tavabi A, Gradhand M, Mavropoulos P, Lotsch B, Dunin-Borkowski
  R, Mokrousov Y, Blügel S and Kläui M 2022 Skyrmionic spin structures in
  layered $\mathrm{Fe_{5}GeTe_{2}}$ up to room temperature

\bibitem{Sassa_DMFT}
Zhu J~X, Janoschek M, Chaves D~S, Cezar J~C, Durakiewicz T, Ronning F, Sassa Y,
  Mansson M, Scott B~L, Wakeham N, Bauer E~D and Thompson J~D 2016 {\em Phys.
  Rev. B\/} {\bf 93}(14) 144404

\bibitem{Sci_adv_FGT_family}
Seo J, Kim D~Y, An E~S, Kim K, Kim G~Y, Hwang S~Y, Kim D~W, Jang B~G, Kim H,
  Eom G, Seo S~Y, Stania R, Muntwiler M, Lee J, Watanabe K, Taniguchi T, Jo
  Y~J, Lee J, Min B~I, Jo M~H, Yeom H~W, Choi S~Y, Shim J~H and Kim J~S 2020
  {\em Sci. Adv.\/} {\bf 6} 8912

\bibitem{SE_jpcl}
Ershadrad S, Ghosh S, Wang D, Kvashnin Y and Sanyal B 2022 {\em J. Phys. Chem.
  Lett.\/} {\bf 13} 4877--4883

\bibitem{Kim_2021}
Kim D, Lee C, Jang B~G, Kim K and Shim J~H 2021 {\em Sci. Rep.\/} {\bf 11}(1)
  17567

\bibitem{may2020tuning}
May A~F, Du M~H, Cooper V~R and McGuire M~A 2020 {\em Phys. Rev. Mater.\/} {\bf
  4} 074008

\bibitem{CLR}
Cococcioni M and de~Gironcoli S 2005 {\em Phys. Rev. B\/} {\bf 71}(3) 035105

\bibitem{F3GT_PRB_2021}
Shen Z~X, Bo X, Cao K, Wan X and He L 2021 {\em Phys. Rev. B\/} {\bf 103}(8)
  085102

\bibitem{May_F3GT_2016}
May A~F, Calder S, Cantoni C, Cao H and McGuire M~A 2016 {\em Phys. Rev. B\/}
  {\bf 93}(1) 014411

\bibitem{Inorg_F3GT_2015}
Verchenko V~Y, Tsirlin A~A, Sobolev A~V, Presniakov I~A and Shevelkov A~V 2015
  {\em Inorg. Chem.\/} {\bf 54}(17) 8598--8607

\bibitem{Japan_F3GT_2013}
Chen B, Yang J, Wang H, Imai M, Ohta H, Michioka C, Yoshimura K and Fang M 2013
  {\em Journal of the Physical Society of Japan\/} {\bf 82} 124711

\bibitem{Riberio_npj2020}
Ribeiro M, Gentile G, Marty A, Dosenovic D, Okuno H, Vergnaud C, Jacquot J~F,
  Jalabert D, Longo D, Ohresser P, Hallal A, Chshiev M, Boulle O, Bonell F and
  Jamet M 2022 {\em npj 2D Mater. and Appl.\/} {\bf 6}(1) 10

\bibitem{Nature_F3GT_2018}
Deng Y, Yu Y, Song Y, Zhang J, Wang N~Z, Sun Z, Yi Y, Wu Y~Z, Wu S, Zhu J, Wang
  J, Chen X~H and Zhang Y 2018 {\em Nature\/} {\bf 563}(67729) 94--99

\bibitem{Georges_DMFT}
Georges A, Kotliar G, Krauth W and Rozenberg M~J 1996 {\em Rev. Mod. Phys.\/}
  {\bf 68}(1) 13--125

\bibitem{SPTF_PRB}
Pourovskii L~V, Katsnelson M~I and Lichtenstein A~I 2005 {\em Phys. Rev. B\/}
  {\bf 72}(11) 115106

\bibitem{Wills2000}
Wills J, Eriksson O, Alouani M and Price D 2000 {\em Full-Potential LMTO Total
  Energy and Force Calculations\/} (Berlin, Heidelberg: Springer Berlin
  Heidelberg)

\bibitem{DMFT_1996}
Georges A, Kotliar G, Krauth W and Rozenberg M~J 1996 {\em Rev. Mod. Phys.\/}
  {\bf 68}(1) 13--125

\bibitem{Volodymyr_2012}
Turkowski V, Kabir A, Nayyar N and Rahman T~S 2012 {\em J. Chem. Phys.\/} {\bf
  136} 114108

\bibitem{Kim_2020}
Kim T~J, Ryee S, Han M~J and Choi S 2020 {\em 2D Materials\/} {\bf 7} 035023

\bibitem{Zhou_2021}
Zhou Z, Pandey S~K and Feng J 2021 {\em Phys. Rev. B\/} {\bf 103}(3) 035137

\bibitem{Kvashnin_2022}
Kvashnin Y~O, Rudenko A~N, Thunstr\"om P, R{\"o}sner M and Katsnelson M~I 2022
  {\em Phys. Rev. B\/} {\bf 105}(20) 205124

\bibitem{Hu_ACS_2020}
Hu X, Zhao Y, Shen X, Krasheninnikov A~V and Chen Zhongfang~andSun L 2020 {\em
  ACS Appl. Mater. Interfaces\/} {\bf 12}(23) 26367--26373

\bibitem{Kondo_Nanolett}
Zhao M, Chen B~B, Xi Y, Zhao Y, Xu H, Zhang H, Cheng N, Feng H, Zhuang J, Pan
  F, Xu X, Hao W, Li W, Zhou S, Dou S~X and Du Y 2021 {\em Nano Lett.\/} {\bf
  21}(14) 6117--6123

\bibitem{Sciadv_F3GT_2018}
Zhang Y, Lu H, Zhu X, Tan S, Feng W, Liu Q, Zhang W, Chen Q, Liu Y, Luo X, Xie
  D, Luo L, Zhang Z and Lai X 2018 {\em Sci. Adv.\/} {\bf 4} eaao6791

\bibitem{Yang2021}
Yang X, Zhou X, Feng W and Yao Y 2021 {\em Phys. Rev. B\/} {\bf 104}(10) 104427

\bibitem{Li_2020}
Li Z, Xia W, Su H, Yu Z, Fu Y, Chen L, Wang X, Yu N, Zou Z and Guo Y 2020 {\em
  Sci. Reports\/} {\bf 10}(1) 15345

\bibitem{May_Nat_Mater_2018}
Fei Z, Huang B, Malinowski P, Wang W, Song T, Sanchez J, Yao W, Xiao D, Zhu X,
  May A~F, Wu W, Cobden D~H, Chu J~H and Xu X 2018 {\em Nat. Mater.\/} {\bf
  17}(9) 778--782

\bibitem{Parkin_2022}
Chakraborty A, Srivastava A~K, Sharma A~K, Gopi A~K, Mohseni K, Ernst A, Deniz
  H, Hazra B~K, Das S, Sessi P, Kostanovskiy I, Ma T, Meyerheim H~L and Parkin
  S~S~P 2022 {\em Adv. Mater.\/} {\bf 34} 2108637

\bibitem{Geetha_2021}
Mayoh D~A, Wood G~D~A, Holt S~J~R, Beckett G, Dekker E~J~L, Lees M~R and
  Balakrishnan G 2021 {\em Cryst. Growth Des.\/} {\bf 21}(12) 6786--6792

\bibitem{Yaroslav_2015}
Kvashnin Y~O, Gr\aa{}n\"as O, Di~Marco I, Katsnelson M~I, Lichtenstein A~I and
  Eriksson O 2015 {\em Phys. Rev. B\/} {\bf 91}(12) 125133

\bibitem{Vladi_PRM_2022}
Borisov V, Xu Q, Ntallis N, Clulow R, Shtender V, Cedervall J, Sahlberg M,
  Wikfeldt K~T, Thonig D, Pereiro M, Bergman A, Delin A and Eriksson O 2022
  {\em Phys. Rev. Materials\/} {\bf 6}(8) 084401

\bibitem{LIECHTENSTEIN198765}
 1987 {\em Journal of Magnetism and Magnetic Materials\/} {\bf 67} 65--74

\bibitem{Yi_2016}
Yi J, Zhuang H, Zou Q, Wu Z, Cao G, Tang S, Calder S~A, Kent P~R~C, Mandrus D
  and Gai Z 2016 {\em 2D Mater.\/} {\bf 4} 011005

\bibitem{Kim_2019}
Kim D, Park S, Lee J, Yoon J, Joo S, Kim T, joon Min K, Park S~Y, Kim C, Moon
  K~W, Lee C, Hong J and Hwang C 2019 {\em Nanotechnology\/} {\bf 30} 245701

\bibitem{Nanoscale_2020}
Jang S~W, Yoon H, Jeong M~Y, Ryee S, Kim H~S and Han M~J 2020 {\em Nanoscale\/}
  {\bf 12}(25) 13501--13506

\bibitem{npj_F3GT}
Roemer R, Liu C and Zou K 2020 {\em npj 2D Mater. Appl.\/} {\bf 4}(1)
  2397--7132

\bibitem{ZHU2021}
Zhu M, You Y, Xu G, Tang J, Gong Y and Xu F 2021 {\em Intermetallics\/} {\bf
  131} 107085

\bibitem{Kresse1999}
Kresse G and Joubert D 1999 {\em Phys. Rev. B\/} {\bf 59} 1758

\bibitem{Kresse1994}
Kresse G and Hafner J 1994 {\em J. Phys. Condens. Matter\/} {\bf 6} 8245

\bibitem{Perdew1996}
Perdew J~P, Burke K and Ernzerhof M 1996 {\em Phys. Rev. Lett.\/} {\bf 77} 3865

\bibitem{Monkhorst1976}
Monkhorst H~J and Pack J~D 1976 {\em Phys. Rev. B\/} {\bf 13} 5188

\bibitem{Wills1987}
Wills J~M and Cooper B~R 1987 {\em Phys. Rev. B\/} {\bf 36}(7) 3809--3823

\bibitem{Oscar}
Grånäs O, {Di Marco} I, Thunström P, Nordström L, Eriksson O, Björkman T
  and Wills J 2012 {\em Comput. Mater. Sci.\/} {\bf 55} 295--302 ISSN 0927-0256

\bibitem{Eriksson_PRB_2007}
Grechnev A, Di~Marco I, Katsnelson M~I, Lichtenstein A~I, Wills J and Eriksson
  O 2007 {\em Phys. Rev. B\/} {\bf 76}(3) 035107

\bibitem{Lichenstein_2005}
Pourovskii L~V, Katsnelson M~I and Lichtenstein A~I 2005 {\em Phys. Rev. B\/}
  {\bf 72}(11) 115106

\bibitem{Kotilar_PRL_2001}
Lichtenstein A~I, Katsnelson M~I and Kotliar G 2001 {\em Phys. Rev. Lett.\/}
  {\bf 87}(6) 067205

\bibitem{Kvashnin2020}
Kvashnin Y~O, Bergman A, Lichtenstein A~I and Katsnelson M~I 2020 {\em Phys.
  Rev. B\/} {\bf 102}(11) 115162

\bibitem{Vladislav_PRB_2021}
Borisov V, Kvashnin Y~O, Ntallis N, Thonig D, Thunstr\"om P, Pereiro M, Bergman
  A, Sj\"oqvist E, Delin A, Nordstr\"om L and Eriksson O 2021 {\em Phys. Rev.
  B\/} {\bf 103}(17) 174422

\bibitem{eriksson2017atomistic}
Eriksson O, Bergman A, Bergqvist L and Hellsvik J 2017 {\em Atomistic Spin
  Dynamics: Foundations and Applications\/} (Oxford university press)

\bibitem{Dudarev}
Dudarev S~L, Botton G~A, Savrasov S~Y, Humphreys C~J and Sutton A~P 1998 {\em
  Phys. Rev. B\/} {\bf 57}(3) 1505--1509

\end{thebibliography}


\end{document}